# Optimal Aggregation Algorithms for Middleware

Ronald Fagin*
Amnon Lotem†
Moni Naor‡

**Abstract:** Assume that each object in a database has $m$ grades, or scores, one for each of $m$ attributes. For example, an object can have a color grade, that tells how red it is, and a shape grade, that tells how round it is. For each attribute, there is a sorted list, which lists each object and its grade under that attribute, sorted by grade (highest grade first). Each object is assigned an overall grade, that is obtained by combining the attribute grades using a fixed monotone *aggregation function*, or *combining rule*, such as min or average.

To determine the top $k$ objects, that is, $k$ objects with the highest overall grades, the naive algorithm must access every object in the database, to find its grade under each attribute. Fagin has given an algorithm ("Fagin's Algorithm", or FA) that is much more efficient. For some monotone aggregation functions, FA is optimal with high probability in the worst case.

We analyze an elegant and remarkably simple algorithm ("the threshold algorithm", or TA) that is optimal in a much stronger sense than FA. We show that TA is essentially optimal, not just for some monotone aggregation functions, but for all of them, and not just in a high-probability worst-case sense, but over every database. Unlike FA, which requires large buffers (whose size may grow unboundedly as the database size grows), TA requires only a small, constant-size buffer. TA allows early stopping, which yields, in a precise sense, an approximate version of the top $k$ answers.

We distinguish two types of access: sorted access (where the middleware system obtains the grade of an object in some sorted list by proceeding through the list sequentially from the top), and random access (where the middleware system requests the grade of object in a list, and obtains it in one step). We consider the scenarios where random access is either impossible, or expensive relative to sorted access, and provide algorithms that are essentially optimal for these cases as well.

*IBM Almaden Research Center, 650 Harry Road, San Jose, California 95120. Email: fagin@almaden.ibm.com
†University of Maryland–College Park, Dept. of Computer Science, College Park, Maryland 20742. Email: lotem@cs.umd.edu
‡Dept. of Computer Science and Applied Mathematics, Weizmann Institute of Science, Rehovot 76100, Israel. Email: naor@wisdom.weizmann.ac.il. The work of this author was performed while a Visiting Scientist at the IBM Almaden Research Center.

# 1  Introduction

Early database systems were required to store only small character strings, such as the entries in a tuple in a traditional relational database. Thus, the data was quite homogeneous. Today, we wish for our database systems to be able to deal not only with character strings (both small and large), but also with a heterogeneous variety of multimedia data (such as images, video, and audio). Furthermore, the data that we wish to access and combine may reside in a variety of data repositories, and we may want our database system to serve as middleware that can access such data.

One fundamental difference between small character strings and multimedia data is that multimedia data may have attributes that are inherently fuzzy. For example, we do not say that a given image is simply either "red" or "not red". Instead, there is a degree of redness, which ranges between 0 (not at all red) and 1 (totally red).

One approach [?] to deal with such fuzzy data is to make use of an *aggregation function* $t$. If $x_1, \ldots, x_m$ (each in the interval $[0, 1]$) are the grades of object $R$ under the $m$ attributes, then $t(x_1, \ldots, x_m)$ is the (overall) grade of object $R$. We shall often abuse notation and write $t(R)$ for the grade $t(x_1, \ldots, x_m)$ of $R$. As we shall discuss, such aggregation functions are useful in other contexts as well. There is a large literature on choices for the aggregation function (see Zimmermann's textbook [?] and the discussion in [?]).

One popular choice for the aggregation function is $\min$. In fact, under the standard rules of fuzzy logic [?], if object $R$ has grade $x_1$ under attribute $A_1$ and $x_2$ under attribute $A_2$, then the grade under the fuzzy conjunction $A_1 \wedge A_2$ is $\min(x_1, x_2)$. Another popular aggregation function is the average (or the sum, in contexts where we do not care if the resulting overall grade no longer lies in the interval $[0, 1]$).

We say that an aggregation function $t$ is *monotone* if $t(x_1, \ldots, x_m) \leq t(x'_1, \ldots, x'_m)$ whenever $x_i \leq x'_i$ for every $i$. Certainly monotonicity is a reasonable property to demand of an aggregation function: if for every attribute, the grade of object $R'$ is at least as high as that of object $R$, then we would expect the overall grade of $R'$ to be at least as high as that of $R$.

The notion of a query is different in a multimedia database system than in a traditional database system. Given a query in a traditional database system (such as a relational database system), there is an unordered set of answers.[1] By contrast, in a multimedia database system, the answer to a query is a "graded" (or "fuzzy") set [?]. A graded set is a set of pairs $(x, g)$, where $x$ is an object, and $g$ (the grade) is a real number in the interval $[0, 1]$. Graded sets are usually presented in sorted order, sorted by grade. As in [?], we shall identify a query with a choice of the aggregation function $t$. The user is typically interested in finding the *top $k$ answers*, where $k$ is a given parameter (such as $k = 1$, $k = 10$, or $k = 100$). This means that we want to obtain $k$ objects (which we may refer to as the "top $k$ objects") with the highest grades on this query, each along with its grade (ties are broken arbitrarily). For convenience, throughout this paper we will think of $k$ as a constant value, and we will consider algorithms for obtaining the top $k$ answers in databases that contain at least $k$ objects.

**Other applications:** There are other applications besides multimedia databases where we make use of an aggregation function to combine grades, and where we want to find the top $k$ answers. One important example is information retrieval [?], where the objects $R$ of interest are documents, the $m$ attributes are search terms $s_1, \ldots, s_m$, and the grade $x_i$ measures the relevance of document $R$ for search term $s_i$, for $1 \leq i \leq m$. It is common to take the aggregation function $t$ to be the sum. That is, the total relevance score of document $R$ when the query consists of the search terms $s_1, \ldots, s_m$ is taken to be $t(x_1, \ldots, x_m) = x_1 + \cdots + x_m$.

---

[1] Of course, in a relational database, the result to a query may be sorted in some way for convenience in presentation, such as sorting department members by salary, but logically speaking, the result is still simply a set, with a crisply-defined collection of members.

Another application arises in a paper by Aksoy and Franklin [?] on scheduling large-scale on-demand data broadcast. In this case each object is a page, and there are two fields. The first field represents the amount of time waited by the earliest user requesting a page, and the second field represents the number of users requesting a page. They make use of the product function $t$ with $t(x_1, x_2) = x_1 x_2$, and they wish to broadcast next the page with the top score.

**The model:** We assume that each database consists of a finite set of *objects*. We shall typically take $N$ to represent the number of objects. Associated with each object $R$ are $m$ *fields* $x_1, \ldots, x_m$, where $x_i \in [0, 1]$ for each $i$. We may refer to $x_i$ as the $i$th field of $R$. The database can be thought of as consisting of a single relation, where one column corresponds to the object id, and the other columns correspond to $m$ attributes of the object. Alternatively, the way we shall think of a database in this paper is as consisting of $m$ sorted lists $L_1, \ldots, L_m$, each of length $N$ (there is one entry in each list for each of the $N$ objects). We may refer to $L_i$ as *list i*. Each entry of $L_i$ is of the form $(R, x_i)$, where $x_i$ is the $i$th field of $R$. Each list $L_i$ is sorted in descending order by the $x_i$ value. We take this simple view of a database, since this view is all that is relevant, as far as our algorithms are concerned. We are taking into account only access costs, and ignoring internal computation costs. Thus, in practice it might well be expensive to compute the field values, but we ignore this issue here, and take the field values as being given.

We consider two modes of access to data. The first mode of access is sorted (or sequential) access. Here the middleware system obtains the grade of an object in one of the sorted lists by proceeding through the list sequentially from the top. Thus, if object $R$ has the $\ell$th highest grade in the $i$th list, then $\ell$ sorted accesses to the $i$th list are required to see this grade under sorted access. The second mode of access is random access. Here, the middleware system requests the grade of object $R$ in the $i$th list, and obtains it in one random access. If there are $s$ sorted accesses and $r$ random accesses, then the *sorted access cost* is $sc_S$, the *random access cost* is $rc_R$, and the *middleware cost* is $sc_S + rc_R$ (the sum of the sorted access cost and the random access cost), for some positive constants $c_S$ and $c_R$.

**Algorithms:** There is an obvious naive algorithm for obtaining the top $k$ answers. Under sorted access, it looks at every entry in each of the $m$ sorted lists, computes (using $t$) the overall grade of every object, and returns the top $k$ answers. The naive algorithm has linear middleware cost (linear in the database size), and thus is not efficient for a large database.

Fagin [?] introduced an algorithm ("Fagin's Algorithm", or FA), which often does much better than the naive algorithm. In the case where the orderings in the sorted lists are probabilistically independent, FA finds the top $k$ answers, over a database with $N$ objects, with middleware cost $O(N^{(m-1)/m} k^{1/m})$, with arbitrarily high probability.[2] Fagin also proved that under this independence assumption, along with an assumption on the aggregation function, every correct algorithm must, with high probability, incur a similar middleware cost in the worst case.

We shall present the "threshold algorithm", or TA. This algorithm was discovered independently by (at least) three groups, including Nepal and Ramakrishna [?] (who were the first to publish), Güntzer, Balke, and Kiessling [?], and ourselves.[3] For more information and comparison, see Section 10 on

---

[2]We shall not discuss the probability model here, including the notion of "independence", since it is off track. For details, see [?].

[3]Our second author first defined TA, and did extensive simulations comparing it to FA, as a project in a database course

related work.

We shall show that TA is optimal in a much stronger sense than FA. We now define this notion of optimality, which we consider to be interesting in its own right.

**Instance optimality:** Let **A** be a class of algorithms, let **D** be a class of databases, and let $cost(\mathcal{A}, \mathcal{D})$ be the middleware cost incurred by running algorithm $\mathcal{A}$ over database $\mathcal{D}$. We say that an algorithm $\mathcal{B}$ is *instance optimal over* **A** *and* **D** if $\mathcal{B} \in \mathbf{A}$ and if for every $\mathcal{A} \in \mathbf{A}$ and every $\mathcal{D} \in \mathbf{D}$ we have

$$cost(\mathcal{B}, \mathcal{D}) = O(cost(\mathcal{A}, \mathcal{D})). \tag{1}$$

Equation (1) means that there are constants $c$ and $c'$ such that $cost(\mathcal{B}, \mathcal{D}) \leq c \cdot cost(\mathcal{A}, \mathcal{D}) + c'$ for every choice of $\mathcal{A} \in \mathbf{A}$ and $\mathcal{D} \in \mathbf{D}$. We refer to $c$ as the *optimality ratio*. Intuitively, instance optimality corresponds to optimality in every instance, as opposed to just the worst case or the average case. FA is optimal in a high-probability worst-case sense under certain assumptions. TA is optimal in a much stronger sense, and without any underlying probabilistic model or probabilistic assumptions: it is instance optimal, for several natural choices of **A** and **D**. In particular, instance optimality holds when **A** is taken to be the class of algorithms that would normally be implemented in practice (since the only algorithms that are excluded are those that make very lucky guesses), and when **D** is taken to be the class of all databases. Instance optimality of TA holds in this case for all monotone aggregation functions. By contrast, high-probability worst-case optimality of FA holds only under the assumption of "strictness" (we shall define strictness later; intuitively, it means that the aggregation function is representing some notion of conjunction).

**Approximation and early stopping:** There are times when the user may be satisfied with an *approximate* top $k$ list. Assume $\theta > 1$. Define a $\theta$-*approximation to the top $k$ answers* for the aggregation function $t$ to be a collection of $k$ objects (each along with its grade) such that for each $y$ among these $k$ objects and each $z$ not among these $k$ objects, $\theta t(y) \geq t(z)$. Note that the same definition with $\theta = 1$ gives the top $k$ answers. We show how to modify TA to give such a $\theta$-approximation (and prove the instance optimality of this modified algorithm under certain assumptions). In fact, we can easily modify TA into an interactive process where at all times the system can show the user its current view of the top $k$ list along with a guarantee about the degree $\theta$ of approximation to the correct answer. At any time, the user can decide, based on this guarantee, whether he would like to stop the process.

**Restricting random access:** As we shall discuss in Section 2, there are some systems where random access is impossible. To deal with such situations, we show in Section 8.1 how to modify TA to obtain an algorithm NRA ("no random accesses") that does no random accesses. We prove that NRA is instance optimal over all algorithms that do not make random accesses and over all databases.

What about situations where random access is not impossible, but simply expensive? Wimmers et al. [?] discuss a number of systems issues that can cause random access to be expensive. Although TA is instance optimal, the optimality ratio depends on the ratio $c_R/c_S$ of the cost of a single random access to the cost of a single sorted access. We define another algorithm that is a combination of TA and NRA, and call it CA ("combined algorithm"). The definition of the algorithm depends on $c_R/c_S$. The motivation is to obtain an algorithm that is not only instance optimal, but whose optimality ratio is independent of $c_R/c_S$. Our original hope was that CA would be instance optimal (with optimality ratio independent of $c_R/c_S$) in those scenarios where TA is instance optimal. Not only does this hope fail, but interestingly enough, we prove that there does not exist any deterministic algorithm, or even taught by Michael Franklin at the University of Maryland–College Park, in the Fall of 1997.

probabilistic algorithm that does not make a mistake, with optimality ratio independent of $c_R/c_S$ in these scenarios! However, we find a new natural scenario where CA is instance optimal, with optimality ratio independent of $c_R/c_S$.

**Outline of paper:** In Section 2, we discuss modes of access (sorted and random) to data. In Section 3, we present FA (Fagin's Algorithm) and its properties. In Section 4, we present TA (the Threshold Algorithm). In Section 5, we define instance optimality, and compare it with related notions, such as competitiveness. In Section 6, we show that TA is instance optimal in several natural scenarios. In the most important scenario, we show that the optimality ratio of TA is best possible. In Section 6.1, we discuss the dependence of the optimality ratio on various parameters. In Section 6.2, we show how to turn TA into an approximation algorithm, and prove instance optimality among approximation algorithms. We also show how the user can prematurely halt TA and in a precise sense, treat its current view of the top $k$ answers as an approximate answer. In Section 7, we consider situations (suggested by Bruno, Gravano, and Marian [**?**]) where sorted access is impossible for certain of the sorted lists. In Section 8, we focus on situations where random accesses are either impossible or expensive. In Section 8.1 we present NRA (No Random Access algorithm), and show its instance optimality among algorithms that make no random accesses. Further, we show that the optimality ratio of NRA is best possible. In Section 8.2 we present CA (Combined Algorithm), which is a result of combining TA and NRA in order to obtain an algorithm that, intuitively, minimizes random accesses. In Section 8.3, we show instance optimality of CA, with an optimality ratio independent of $c_R/c_S$, in a natural scenario. In Section 8.4, we show that the careful choice made by CA of which random accesses to make is necessary for instance optimality with an optimality ratio independent of $c_R/c_S$. We also compare and contrast CA versus TA. In Section 9, we prove various lower bounds on the optimality ratio, both for deterministic algorithms and for probabilistic algorithms that never make a mistake. We summarize our upper and lower bounds in Section 9.1. In Section 10 we discuss related work. In Section 11, we give our conclusions, and state some open problems.

## 2 Modes of Access to Data

Issues of efficient query evaluation in a middleware system are very different from those in a traditional database system. This is because the middleware system receives answers to queries from various subsystems, which can be accessed only in limited ways. What do we assume about the interface between a middleware system and a subsystem? Let us consider QBIC[4] [**?**] ("Query By Image Content") as a subsystem. QBIC can search for images by various visual characteristics such as color and texture (and an experimental version can search also by shape). In response to a query, such as *Color='red'*, the subsystem will output the graded set consisting of all objects, one by one, each along with its grade under the query, in sorted order based on grade, until the middleware system tells the subsystem to halt. Then the middleware system could later tell the subsystem to resume outputting the graded set where it left off. Alternatively, the middleware system could ask the subsystem for, say, the top 10 objects in sorted order, each along with its grade. then request the next 10, etc. In both cases, this corresponds to what we have referred to as "sorted access".

There is another way that we might expect the middleware system to interact with the subsystem. Specifically, the middleware system might ask the subsystem for the grade (with respect to a query) of any given object. This corresponds to what we have referred to as "random access". In fact, QBIC allows both sorted and random access.

---
[4]QBIC is a trademark of IBM Corporation.

There are some situations where the middleware system is not allowed random access to some subsystem. An example might occur when the middleware system is a text retrieval system, and the subsystems are search engines. Thus, there does not seem to be a way to ask a major search engine on the web for its internal score on some document of our choice under a query.

Our measure of cost corresponds intuitively to the cost incurred by the middleware system in processing information passed to it from a subsystem such as QBIC. As before, if there are $s$ sorted accesses and $r$ random accesses, then the *middleware cost* is taken to be $sc_S + rc_R$, for some positive constants $c_S$ and $c_R$. The fact that $c_S$ and $c_R$ may be different reflects the fact that the cost to a middleware system of a sorted access and of a random access may be different.

## 3 Fagin's Algorithm

In this section, we discuss FA (Fagin's Algorithm) [?].This algorithm is implemented in Garlic [?], an experimental IBM middleware system; see [?] for interesting details about the implementation and performance in practice. Chaudhuri and Gravano [?] consider ways to simulate FA by using "filter conditions", which might say, for example, that the color score is at least 0.2.[5] FA works as follows.

1. Do sorted access in parallel to each of the $m$ sorted lists $L_i$. (By "in parallel", we mean that we access the top member of each of the lists under sorted access, then we access the second member of each of the lists, and so on.)[6] Wait until there are at least $k$ "matches", that is, wait until there is a set $H$ of at least $k$ objects such that each of these objects has been seen in each of the $m$ lists.

2. For each object $R$ that has been seen, do random access as needed to each of the lists $L_i$ to find the $i$th field $x_i$ of $R$.

3. Compute the grade $t(R) = t(x_1, \ldots, x_m)$ for each object $R$ that has been seen. Let $Y$ be a set containing the $k$ objects that have been seen with the highest grades (ties are broken arbitrarily). The output is then the graded set $\{(R, t(R)) \mid R \in Y\}$.

It is fairly easy to show [?] that this algorithm is correct for monotone aggregation functions $t$ (that is, that the algorithm successfully finds the top $k$ answers). If there are $N$ objects in the database, and if the orderings in the sorted lists are probabilistically independent, then the middleware cost of FA is $O(N^{(m-1)/m} k^{1/m})$, with arbitrarily high probability [?].

An aggregation function $t$ is *strict* [?] if $t(x_1, \ldots, x_m) = 1$ holds precisely when $x_i = 1$ for every $i$. Thus, an aggregation function is strict if it takes on the maximal value of 1 precisely when each argument takes on this maximal value. We would certainly expect an aggregation function representing the conjunction to be strict (see the discussion in [?]). In fact, it is reasonable to think of strictness as being a key characterizing feature of the conjunction.

---

[5]Chaudhuri and Gravano originally saw an early version of the conference paper (in the 1996 ACM Symposium on Principles of Database Systems) that expanded into the journal version [?].

[6]It is not actually important that the lists be accessed "in lockstep". In practice, it may be convenient to allow the sorted lists to be accessed at different rates, in batches, etc. Each of the algorithms in this paper where there is "sorted access in parallel" remain correct even when sorted access is not in lockstep. Furthermore, all of our instance optimality results continue to hold even when sorted access is not in lockstep, as long as the rates of sorted access of the lists are within constant multiples of each other.

Fagin shows that his algorithm is optimal with high probability in the worst case if the aggregation function is strict (so that, intuitively, we are dealing with a notion of conjunction), and if the orderings in the sorted lists are probabilistically independent. In fact, the access pattern of FA is oblivious to the choice of aggregation function, and so for each fixed database, the middleware cost of FA is exactly the same no matter what the aggregation function is. This is true even for a constant aggregation function; in this case, of course, there is a trivial algorithm that gives us the top $k$ answers (any $k$ objects will do) with $O(1)$ middleware cost. So FA is not optimal in any sense for some monotone aggregation functions $t$. As a more interesting example, when the aggregation function is max (which is not strict), it is shown in [?] that there is a simple algorithm that makes at most $mk$ sorted accesses and no random accesses that finds the top $k$ answers. By contrast, as we shall see, the algorithm TA is instance optimal for every monotone aggregation function, under very weak assumptions.

Even in the cases where FA is optimal, this optimality holds only in the worst case, with high probability. This leaves open the possibility that there are some algorithms that have much better middleware cost than FA over certain databases. The algorithm TA, which we now discuss, is such an algorithm.

## 4 The Threshold Algorithm

We now present the threshold algorithm (TA).

1. Do sorted access in parallel to each of the $m$ sorted lists $L_i$. As an object $R$ is seen under sorted access in some list, do random access to the other lists to find the grade $x_i$ of object $R$ in every list $L_i$.[7] Then compute the grade $t(R) = t(x_1, \ldots, x_m)$ of object $R$. If this grade is one of the $k$ highest we have seen, then remember object $R$ and its grade $t(R)$ (ties are broken arbitrarily, so that only $k$ objects and their grades need to be remembered at any time).

2. For each list $L_i$, let $\underline{x}_i$ be the grade of the last object seen under sorted access. Define the *threshold value* $\tau$ to be $t(\underline{x}_1, \ldots, \underline{x}_m)$. As soon as at least $k$ objects have been seen whose grade is at least equal to $\tau$, then halt.

3. Let $Y$ be a set containing the $k$ objects that have been seen with the highest grades. The output is then the graded set $\{(R, t(R)) \mid R \in Y\}$.

We now show that TA is correct for each monotone aggregation function $t$.

**Theorem 4.1:** *If the aggregation function $t$ is monotone, then TA correctly finds the top $k$ answers.*

**Proof:** Let $Y$ be as in Step 3 of TA. We need only show that every member of $Y$ has at least as high a grade as every object $z$ not in $Y$. By definition of $Y$, this is the case for each object $z$ that has been seen in running TA. So assume that $z$ was not seen. Assume that the fields of $z$ are $x_1, \ldots, x_m$. Therefore, $x_i \leq \underline{x}_i$, for every $i$. Hence, $t(z) = t(x_1, \ldots, x_m) \leq t(\underline{x}_1, \ldots, \underline{x}_m) = \tau$, where the inequality follows by monotonicity of $t$. But by definition of $Y$, for every $y$ in $Y$ we have $t(y) \geq \tau$. Therefore, for every $y$ in $Y$ we have $t(y) \geq \tau \geq t(z)$, as desired. □

---

[7]It may seem wasteful to do random access to find a grade that was already determined earlier. As we discuss later, this is done in order to avoid unbounded buffers.

We now show that the stopping rule for TA always occurs at least as early as the stopping rule for FA (that is, with no more sorted accesses than FA). In FA, if $R$ is an object that has appeared under sorted access in every list, then by monotonicity, the grade of $R$ is at least equal to the threshold value. Therefore, when there are at least $k$ objects, each of which has appeared under sorted access in every list (the stopping rule for FA), there are at least $k$ objects whose grade is at least equal to the threshold value (the stopping rule for TA).

This implies that for every database, the sorted access cost for TA is at most that of FA. This does not imply that the middleware cost for TA is always at most that of FA, since TA may do more random accesses than FA. However, since the middleware cost of TA is at most the sorted access cost times a constant (independent of the database size), it does follow that the middleware cost of TA is at most a constant times that of FA. In fact, we shall show that TA is instance optimal, under natural assumptions.

We now consider the intuition behind TA. For simplicity, we discuss first the case where $k = 1$, that is, where the user is trying to determine the top answer. Assume that we are at a stage in the algorithm where we have not yet seen any object whose (overall) grade is at least as big as the threshold value $\tau$. The intuition is that at this point, we do not know the top answer, since the next object we see under sorted access could have overall grade $\tau$, and hence bigger than the grade of any object seen so far. Furthermore, once we do see an object whose grade is at least $\tau$, then it is safe to halt, as we see from the proof of Theorem 4.1. Thus, intuitively, the stopping rule of TA says: "Halt as soon as you know you have seen the top answer." Similarly, for general $k$, the stopping rule of TA says, intuitively, "Halt as soon as you know you have seen the top $k$ answers." So we could consider TA as being an implementation of the following "program":

**Do sorted access (and the corresponding random access) until you know you have seen the top $k$ answers.**

This very high–level "program" is a *knowledge–based program* [**?**]. In fact, TA was designed by thinking in terms of this knowledge-based program. The fact that TA corresponds to this knowledge–based program is what is behind instance optimality of TA.

Later, we shall give other scenarios (situations where random accesses are either impossible or expensive) where we implement the following more general knowledge–based program:

**Gather what information you need to allow you to know the top $k$ answers, and then halt.**

In each of our scenarios, the implementation of this second knowledge-based program is different. When we consider the scenario where random accesses are expensive relative to sorted accesses, but are not impossible, we need an additional design principle to decide how to gather the information, in order to design an instance optimal algorithm.

The next theorem, which follows immediately from the definition of TA, gives a simple but important property of TA that further distinguishes TA from FA.

**Theorem 4.2:** *TA requires only bounded buffers, whose size is independent of the size of the database.*

**Proof:** Other than a little bit of bookkeeping, all that TA must remember is the current top $k$ objects and their grades, and (pointers to) the last objects seen in sorted order in each list. □

By contrast, FA requires buffers that grow arbitrarily large as the database grows, since FA must remember every object it has seen in sorted order in every list, in order to check for matching objects in the various lists.

There is a price to pay for the bounded buffers. Thus, for every time an object is found under sorted access, TA may do $m - 1$ random accesses (where $m$ is the number of lists), to find the grade of the object in the other lists. This is in spite of the fact that this object may have already been seen in these other lists.

## 5  Instance optimality

In order to compare instance optimality with other notions from the literature, we generalize slightly the definition from that given in the introduction. Let **A** be a class of algorithms, and let **D** be a class of legal inputs to the algorithms. We assume that we are considering a particular nonnegative performance cost measure $cost(\mathcal{A}, \mathcal{D})$, which represents the amount of a resource consumed by running the algorithm $\mathcal{A} \in \mathbf{A}$ on input $\mathcal{D} \in \mathbf{D}$. This cost could be the running time of algorithm $\mathcal{A}$ on input $\mathcal{D}$, or in this paper, the middleware cost incurred by running algorithm $\mathcal{A}$ over database $\mathcal{D}$.

We say that an algorithm $\mathcal{B}$ is *instance optimal over* **A** *and* **D** if $\mathcal{B} \in \mathbf{A}$ and if for every $\mathcal{A} \in \mathbf{A}$ and every $\mathcal{D} \in \mathbf{D}$ we have

$$cost(\mathcal{B}, \mathcal{D}) = O(cost(\mathcal{A}, \mathcal{D})). \tag{2}$$

Equation (2) means that there are constants $c$ and $c'$ such that $cost(\mathcal{B}, \mathcal{D}) \leq c \cdot cost(\mathcal{A}, \mathcal{D}) + c'$ for every choice of $\mathcal{A} \in \mathbf{A}$ and $\mathcal{D} \in \mathbf{D}$. We refer to $c$ as the *optimality ratio*. It is similar to the competitive ratio in competitive analysis (we shall discuss competitive analysis shortly). We use the word "optimal" to reflect that fact that $\mathcal{B}$ is essentially the best algorithm in **A**.

Intuitively, instance optimality corresponds to optimality in every instance, as opposed to just the worst case or the average case. There are many algorithms that are optimal in a worst-case sense, but are not instance optimal. An example is binary search: in the worst case, binary search is guaranteed to require no more than $\log N$ probes, for $N$ data items. However, for each instance, a positive answer can be obtained in one probe, and a negative answer in two probes.

We consider a nondeterministic algorithm correct if on no branch does it make a mistake. We take the middleware cost of a nondeterministic algorithm to be the minimal cost over all branches where it halts with the top $k$ answers. We take the middleware cost of a probabilistic algorithm to be the expected cost (over all probabilistic choices by the algorithm). When we say that a deterministic algorithm $\mathcal{B}$ is instance optimal over **A** and **D**, then we are really comparing $\mathcal{B}$ against the best nondeterministic algorithm, even if **A** contains only deterministic algorithms. This is because for each $\mathcal{D} \in \mathbf{D}$, there is always a deterministic algorithm that makes the same choices on $\mathcal{D}$ as the nondeterministic algorithm. We can view the cost of the best nondeterministic algorithm that produces the top $k$ answers over a given database as the cost of the shortest proof for that database that these are really the top $k$ answers. So instance optimality is quite strong: the cost of an instance optimal algorithm is essentially the cost of the shortest proof. Similarly, we can view **A** as if it contains also probabilistic algorithms that never make a mistake. For convenience, in our proofs we shall always assume that **A** contains only deterministic algorithms, since the results carry over automatically to nondeterministic algorithms and to probabilistic algorithms that never make a mistake.

The definition we have given for instance optimality is formally the same definition as is used in *competitive analysis* [**?**, **?**], except that in competitive analysis, (1) we do not assume that $\mathcal{B} \in \mathbf{A}$, and (2) $cost(\mathcal{A}, \mathcal{D})$ does not typically represent a performance cost. In competitive analysis, typically (a) **D** is a class of instances of a particular problem, (b) **A** is the class of offline algorithms that give a solution to the instances in **D**, (c) $cost(\mathcal{A}, \mathcal{D})$ is a number that represents the goodness of the solution (where bigger numbers correspond to a worse solution), and (d) $\mathcal{B}$ is a particular online algorithm. In this case, the online algorithm $\mathcal{B}$ is said to be *competitive*. The intuition is that a competitive online algorithm may perform poorly in some instances, but only on instances where every offline algorithm would also perform poorly.

Another example where the framework of instance optimality appears, but again without the assumption that $\mathcal{B} \in \mathbf{A}$, and again where $cost(\mathcal{A}, \mathcal{D})$ does not represent a performance cost, is in the context of *approximation algorithms* [**?**]. In this case, (a) **D** is a class of instances of a particular problem, (b) **A** is the class of algorithms that solve the instances in **D** exactly (in cases of interest, these algorithms are not polynomial-time algorithms), (c) $cost(\mathcal{A}, \mathcal{D})$ is the value of the resulting answer when algorithm $\mathcal{A}$ is applied to input $\mathcal{D}$, and (d) $\mathcal{B}$ is a particular polynomial-time algorithm.

Dagum et al. [**?**] give an interesting example of what we would call an instance optimal algorithm. They consider the problem of determining the mean of an unknown random variable by Monte Carlo estimation. In their case, (a) **D** is the class of random variables distributed in the interval $[0, 1]$, (b) **A** is the class of algorithms that, by repeatedly doing independent evaluations of a random variable and then averaging the results, obtain an estimate of the mean of the random variable to within a given precision with a given probability, (c) $cost(\mathcal{A}, \mathcal{D})$ is the expected number of independent evaluations of the random variable $\mathcal{D}$ under algorithm $\mathcal{A}$, and (d) $\mathcal{B}$ is their algorithm, which they call $\mathcal{AA}$ for "approximation algorithm". Their main result says, in our terminology, that $\mathcal{AA}$ is instance optimal over **A** and **D**.

Demaine et al. [**?**] give an example of an algorithm that is close to instance optimal. They consider the problem of finding the intersection, union, or difference of a collection of sorted sets. In their case, (a) **D** is the class of instances of collections of sorted sets, (b) **A** is the class of algorithms that do pairwise comparisons among elements, (c) $cost(\mathcal{A}, \mathcal{D})$ is the running time (number of comparisons) in running algorithm $\mathcal{A}$ on instance $\mathcal{D}$, and (d) $\mathcal{B}$ is their algorithm. In a certain sense, their algorithm is close to what we would call instance optimal (to explain the details would take us too far astray).

## 6  Instance Optimality of the Threshold Algorithm

In this section, we investigate the instance optimality of TA. We begin with an intuitive argument that TA is instance optimal. If $\mathcal{A}$ is an algorithm that stops sooner than TA on some database, before $\mathcal{A}$ finds $k$ objects whose grade is at least equal to the threshold value $\tau$, then $\mathcal{A}$ must make a mistake on some database, since the next object in each list might have grade $\underline{x}_i$ in each list $i$, and hence have grade $t(\underline{x}_1, \ldots, \underline{x}_m) = \tau$. This new object, which $\mathcal{A}$ has not even seen, has a higher grade than some object in the top $k$ list that was output by $\mathcal{A}$, and so $\mathcal{A}$ erred by stopping too soon. We would like to convert this intuitive argument into a proof that for every monotone aggregation function, TA is instance optimal over all algorithms that correctly find the top $k$ answers, over the class of all databases. However, as we shall see, the situation is actually somewhat delicate. We first make a distinction between algorithms that "make wild guesses" (that is, perform random access on objects not previously encountered by sorted access) and those that do not. (Neither FA nor TA make wild guesses, nor does any "natural" algorithm in our context.) Our first theorem (Theorem 6.1) says that for every monotone aggregation function, TA is instance optimal over all algorithms that correctly find the top $k$ answers *and that do not make wild guesses*, over the class of *all* databases. We then show that this distinction (wild guesses vs. no wild guesses) is essential: if algorithms that make wild guesses are allowed in the class **A** of algorithms that an instance optimal algorithm must compete against, then *no* algorithm is instance optimal (Example 6.3 and Theorem 6.4). The heart of this example (and the corresponding theorem) is the fact that there may be multiple objects with the same grade in some list. Indeed, once we restrict our attention to databases where no two objects have the same value in the same list, and make a slight, natural additional restriction on the aggregation function beyond monotonicity, then TA is instance optimal over *all* algorithms that correctly find the top $k$ answers (Theorem 6.5).

In Section 6.2 we consider instance optimality in the situation where we relax the problem of finding the top $k$ objects into finding *approximately* the top $k$.

We now give our first positive result on instance optimality of TA. We say that an algorithm *makes wild guesses* if it does random access to find the grade of some object $R$ in some list before the algorithm has seen $R$ under sorted access. That is, an algorithm makes wild guesses if the first grade that it obtains for some object $R$ is under random access. We would not normally implement algorithms that make wild guesses. In fact, there are some contexts where it would not even be possible to make wild guesses (such as a database context where the algorithm could not know the name of an object it has not already seen). However, making a lucky wild guess can help, as we show later (Example 6.3).

We now show instance optimality of TA among algorithms that do not make wild guesses. In this theorem, when we take **D** to be the class of all databases, we really mean that **D** is the class of all databases that involve sorted lists corresponding to the arguments of the aggregation function $t$. We are taking $k$ (where we are trying to find the top $k$ answers) and the aggregation function $t$ to be fixed. Since we are taking $t$ to be fixed, we are thereby taking the number $m$ of arguments of $t$ (that is, the number of sorted lists) to be fixed. In Section 6.1, we discuss the assumptions that $k$ and $m$ are constant.

**Theorem 6.1:** *Assume that the aggregation function $t$ is monotone. Let **D** be the class of all databases. Let **A** be the class of all algorithms that correctly find the top $k$ answers for $t$ for every database and that do not make wild guesses. Then TA is instance optimal over **A** and **D**.*

**Proof:** Assume that $\mathcal{A} \in \mathbf{A}$, and that algorithm $\mathcal{A}$ is run over database $\mathcal{D}$. Assume that algorithm $\mathcal{A}$ halts at depth $d$ (that is, if $d_i$ is the number of objects seen under sorted access to list $i$, for $1 \leq i \leq m$, then $d = \max_i d_i$). Assume that $\mathcal{A}$ sees $a$ distinct objects (some possibly multiple times). In particular, $a \geq d$. Since $\mathcal{A}$ makes no wild guesses, and sees $a$ distinct objects, it must make at least $a$ sorted accesses, and so its middleware cost is at least $ac_S$. We shall show that TA halts on $\mathcal{D}$ by depth $a + k$. Hence, the middleware cost of TA is at most $(a+k)mc_S + (a+k)m(m-1)c_R$, which is $amc_S + am(m-1)c_R$ plus an additive constant of $kmc_S + km(m-1)c_R$. So the optimality ratio of TA is at most $\frac{amc_S + am(m-1)c_R}{ac_S} = m + m(m-1)c_R/c_S$. (Later, we shall show that if the aggregation function is strict, then this is precisely the optimality ratio of TA, and this is best possible.)

Note that for each choice of $d'$, the algorithm TA sees at least $d'$ objects by depth $d'$ (this is because by depth $d'$ it has made $md'$ sorted accesses, and each object is accessed at most $m$ times under sorted access). Let $Y$ be the output set of $\mathcal{A}$ (consisting of the top $k$ objects). If there are at most $k$ objects that $\mathcal{A}$ does not see, then TA halts by depth $a + k$ (after having seen every object), and we are done. So assume that there are at least $k + 1$ objects that $\mathcal{A}$ does not see. Since $Y$ is of size $k$, there is some object $V$ that $\mathcal{A}$ does not see and that is not in $Y$.

Let $\tau_\mathcal{A}$ be the threshold value when algorithm $\mathcal{A}$ halts. This means that if $\underline{x}_i$ is the grade of the last object seen under sorted access to list $i$ for algorithm $\mathcal{A}$, for $1 \leq i \leq m$, then $\tau_\mathcal{A} = t(\underline{x}_1, \ldots, \underline{x}_m)$. (If list $i$ is not accessed under sorted access, we take $\underline{x}_i = 1$.) Let us call an object $R$ *big* if $t(R) \geq \tau_\mathcal{A}$, and otherwise call object $R$ *small*.

We now show that every member $R$ of $Y$ is big. Define a database $\mathcal{D}'$ to be just like $\mathcal{D}$, except that object $V$ has grade $\underline{x}_i$ in the $i$th list, for $1 \leq i \leq m$. Put $V$ in list $i$ below all other objects with grade $\underline{x}_i$ in list $i$ (for $1 \leq i \leq m$). Algorithm $\mathcal{A}$ performs exactly the same, and in particular gives the same output, for databases $\mathcal{D}$ and $\mathcal{D}'$. Therefore, algorithm $\mathcal{A}$ has $R$, but not $V$, in its output for database $\mathcal{D}'$. Since the grade of $V$ in $\mathcal{D}'$ is $\tau_\mathcal{A}$, it follows by correctness of $\mathcal{A}$ that $R$ is big, as desired.

There are now two cases, depending on whether or not algorithm $\mathcal{A}$ sees every member of its output set $Y$.[8]

*Case 1:* Algorithm $\mathcal{A}$ sees every member of $Y$. Then by depth $d$, TA will see every member of $Y$. Since, as we showed, each member of $Y$ is big, it follows that TA halts by depth $d \leq a < a + k$, as desired.

*Case 2:* Algorithm $\mathcal{A}$ does not see some member $R$ of $Y$. We now show that every object $R'$ that is not seen by $\mathcal{A}$ must be big. Define a database $\mathcal{D}'$ that is just like $\mathcal{D}$ on every object seen by $\mathcal{A}$. Let the grade of $V$ in list $i$ be $\underline{x}_i$, and put $V$ in list $i$ below all other objects with grade $\underline{x}_i$ in list $i$ (for $1 \leq i \leq m$). Therefore, the grade of $V$ in database $\mathcal{D}'$ is $\tau_\mathcal{A}$. Since $\mathcal{A}$ cannot distinguish between $\mathcal{D}$ and $\mathcal{D}'$, it has the same output on $\mathcal{D}$ and $\mathcal{D}'$. Since $\mathcal{A}$ does not see $R$ and does not see $R'$, it has no information to distinguish between $R$ and $R'$. Therefore, it must have been able to give $R'$ in its output without making a mistake. But if $R'$ is in the output and not $V$, then by correctness of $\mathcal{A}$, it follows that $R'$ is big. So $R'$ is big, as desired.

Since $\mathcal{A}$ sees $a$ objects, and since TA sees at least $a + k$ objects by depth $a + k$, it follows that by depth $a + k$, TA sees at least $k$ objects not seen by $\mathcal{A}$. We have shown that every object that is not seen by $\mathcal{A}$ is big. Therefore, by depth $a + k$, TA sees at least $k$ big objects. So TA halts by depth $a + k$, as desired. □

The next result is a corollary of the proof of Theorem 6.1 and of a lower bound in Section 9 (all of our results on lower bounds appear in Section 9). Specifically, in the proof of Theorem 6.1, we showed that under the assumptions of Theorem 6.1 (no wild guesses), the optimality ratio of TA is at most $m + m(m - 1)c_R/c_S$. The next result says that if the aggregation function is strict, then the optimality ratio is precisely this value, and this is best possible. Recall that an aggregation function $t$ is strict if $t(x_1, \ldots, x_m) = 1$ holds precisely when $x_i = 1$ for every $i$. Intuitively, strictness means that the aggregation function is representing some notion of conjunction.

**Corollary 6.2:** *Let $t$ be an arbitrary monotone, strict aggregation function with $m$ arguments. Let $\mathbf{D}$ be the class of all databases. Let $\mathbf{A}$ be the class of all algorithms that correctly find the top $k$ answers for $t$ for every database and that do not make wild guesses. Then TA is instance optimal over $\mathbf{A}$ and $\mathbf{D}$, with optimality ratio $m + m(m - 1)c_R/c_S$. No deterministic algorithm has a lower optimality ratio.*

---

[8]For the sake of generality, we are allowing the possibility that algorithm $\mathcal{A}$ can output an object that it has not seen. We discuss this issue more in Section 6.1.

| $L_1$ |
|---|
| $(1,1)$ |
| $(2,1)$ |
| $(3,1)$ |
| ... |
| $(n+1,1)$ |
| $(n+2,0)$ |
| $(n+3,0)$ |
| ... |
| $(2n+1,0)$ |

| $L_2$ |
|---|
| $(2n+1,1)$ |
| $(2n,1)$ |
| $(2n-1,1)$ |
| ... |
| $(n+1,1)$ |
| $(n,0)$ |
| $(n-1,0)$ |
| ... |
| $(1,0)$ |

Figure 1: Database for Example 6.3

**Proof:** In the proof of Theorem 6.1, it is shown that TA has an optimality ratio of at most $m + m(m-1)c_R/c_S$ for an arbitrary monotone aggregation function, The lower bound follows from Theorem 9.1. □

We cannot drop the assumption of strictness in Corollary 6.2. For example, let the aggregation function be max (which is not strict). It is easy to see that TA halts after $k$ rounds of sorted access, and its optimality ratio is $m$ (which, we might add, is best possible for max).[9]

What if we were to consider only the sorted access cost? This corresponds to taking $c_R = 0$. Then we see from Corollary 6.2 that the optimality ratio of TA is $m$. Furthermore, it follows easily from the proof of Theorem 9.1 that if the aggregation function is strict, and if $c_R = 0$, then this is best possible: no deterministic algorithm has a lower optimality ratio than $m$.[10]

What if we were to consider only the random access cost? This corresponds to taking $c_S = 0$. In this case, TA is far from instance optimal. The naive algorithm, which does sorted access to every object in every list, does no random accesses, and so has a sorted access cost of 0.

We now show that making a lucky wild guess can help.

**Example 6.3:** Assume that there are $2n+1$ objects, which we will call simply $1, 2, \ldots, 2n+1$, and there are two lists $L_1$ and $L_2$ (see Figure 1). Assume that in list $L_1$, the objects are in the order $1, 2, \ldots, 2n+1$, where the top $n+1$ objects $1, 2, \ldots, n+1$ all have grade 1, and the remaining $n$ objects $n+2, n+3, \ldots, 2n+1$ all have grade 0. Assume that in list $L_2$, the objects are in the reverse order $2n+1, 2n, \ldots, 1$, where the bottom $n$ objects $1, \ldots, n$ all have grade 0, and the remaining $n+1$ objects $n+1, n+2, \ldots, 2n+1$ all have grade 1. Assume that the aggregation function is min, and that we are interested in finding the top answer (i.e., $k=1$). It is clear that the top answer is object $n+1$ with overall grade 1 (every object except object $n+1$ has overall grade 0).

An algorithm that makes a wild guess and asks for the grade of object $n+1$ in both lists would determine the correct answer and be able to halt safely after two random accesses and no sorted ac-cesses.[11] However, let $\mathcal{A}$ be any algorithm (such as TA) that does not make wild guesses. Since the winning object $n + 1$ is in the middle of both sorted lists, it follows that at least $n + 1$ sorted accesses would be required before algorithm $\mathcal{A}$ would even see the winning object. □

---

[9] Note that the instance optimality of TA, as given by Theorem 6.1, holds whether or not the aggregation function is strict. For example, the instance optimality of TA as given by Theorem 6.1 holds even when the aggregation function is max. This is in contrast to the situation with FA, where high-probability worst-case optimality fails when the aggregation function is max. Corollary 6.2 makes use of the assumption of strictness only in order to show that the optimality ratio of TA is then precisely $m + m(m-1)c_R/c_S$, and that this is best possible.

[10] We are assuming in this paper that $c_R$ and $c_S$ are both strictly positive. However, Corollary 6.2 and the proof of Theorem 9.1 would still hold if we were to allow $c_R$ to be 0.

What if we were to enlarge the class **A** of algorithms to allow queries of the form "Which object has the $i$th largest grade in list $j$, and what is its grade in list $j$?" We then see from Example 6.3, where we replace the wild guess by the query that asks for the object with the $(n + 1)$st largest grade in each list, that TA is not instance optimal. Effectively, these new queries are "just as bad" as wild guesses.

Example 6.3 shows that TA is not instance optimal over the class **A** of all algorithms that find the top answer for min (with two arguments) and the class **D** of all databases. The next theorem says that under these circumstances, not only is TA not instance optimal, but neither is any algorithm.

**Theorem 6.4:** *Let **D** be the class of all databases. Let **A** be the class of all algorithms that correctly find the top answer for min (with two arguments) for every database. There is no deterministic algorithm (or even probabilistic algorithm that never makes a mistake) that is instance optimal over **A** and **D**.*

**Proof:** Let us modify Example 6.3 to obtain a family of databases, each with two sorted lists. The first list has the objects $1, 2, \ldots, 2n + 1$ in some order, with the top $n + 1$ objects having grade 1, and the remaining $n$ objects having grade 0. The second list has the objects in the reverse order, again with the top $n + 1$ objects having grade 1, and the remaining $n$ objects having grade 0. As before, there is a unique object with overall grade 1 (namely, the object in the middle of both orderings), and every remaining object has overall grade 0.

Let $\mathcal{A}$ be an arbitrary deterministic algorithm in **A**. Consider the following distribution on databases: each member is as above, and the ordering of the first list is chosen uniformly at random (with the ordering of the second list the reverse of the ordering of the first list). It is easy to see that the expected number of accesses (sorted and random together) of algorithm $\mathcal{A}$ under this distribution in order to even see the winning object is at least $n+1$. Since there must be some database where the number of accesses is at least equal to the expected number of accesses, the number of accesses on this database is at least $n + 1$. However, as in Example 6.3, there is an algorithm that makes only 2 random accesses and no sorted accesses. Therefore, the optimality ratio can be arbitrarily large. The theorem follows (in the deterministic case).

For probabilistic algorithms that never make a mistake, we appeal to Yao's Minimax Principle [?] (see also [?, Section 2.2], and see [?, Lemma 4] for a simple proof), which says that the expected cost of the optimal deterministic algorithm for an arbitrary input distribution is a lower bound on the expected cost of the optimal probabilistic algorithm that never makes a mistake. □

Although, as we noted earlier, algorithms that make wild guesses would not normally be implemented in practice, it is still interesting to consider them. This is because of our interpretation of instance optimality of an algorithm $\mathcal{A}$ as saying that its cost is essentially the same as the cost of the shortest proof for that database that these are really the top $k$ answers. If we consider algorithms that allow wild guesses, then we are allowing a larger class of proofs. Thus, in Example 6.3, the fact that object $n + 1$ has (overall) grade 1 is a proof that it is the top answer.

---

[11]The algorithm could halt safely, since it "knows" that it has found an object with the maximal possible grade of 1 (this grade is maximal, since we are assuming that all grades lie between 0 and 1). Even if we did not assume that all grades lie between 0 and 1, one sorted access to either list would provide the information that each overall grade in the database is at most 1.

We say that an aggregation function $t$ is *strictly monotone*[12] if $t(x_1, \ldots, x_m) < t(x'_1, \ldots, x'_m)$ whenever $x_i < x'_i$ for every $i$. Although average and min are strictly monotone, there are aggregation functions suggested in the literature for representing conjunction and disjunction that are monotone but not strictly monotone (see [?] and [?] for examples). We say that a database $\mathcal{D}$ *satisfies the distinctness property* if for each $i$, no two objects in $\mathcal{D}$ have the same grade in list $L_i$, that is, if the grades in list $L_i$ are distinct. We now show that these conditions guarantee optimality of TA even among algorithms that make wild guesses.

**Theorem 6.5:** *Assume that the aggregation function $t$ is strictly monotone. Let **D** be the class of all databases that satisfy the distinctness property. Let **A** be the class of all algorithms that correctly find the top $k$ answers for $t$ for every database in **D**. Then TA is instance optimal over **A** and **D**.*

**Proof:** Assume that $\mathcal{A} \in \mathbf{A}$, and that algorithm $\mathcal{A}$ is run over database $\mathcal{D} \in \mathbf{D}$. Assume that $\mathcal{A}$ sees $a$ distinct objects (some possibly multiple times). We shall show that TA halts on $\mathcal{D}$ by depth $a + k$. Hence, TA makes at most $m^2(a+k)$ accesses, which is $m^2 a$ plus an additive constant of $m^2 k$. It follows easily that the optimality ratio of TA is at most $cm^2$, where $c = \max\{c_R/c_S, c_S/c_R\}$.

If there are at most $k$ objects that $\mathcal{A}$ does not see, then TA halts by depth $a + k$ (after having seen every object), and we are done. So assume that there are at least $k + 1$ objects that $\mathcal{A}$ does not see. Since $Y$ is of size $k$, there is some object $V$ that $\mathcal{A}$ does not see and that is not in $Y$. We shall show that TA halts on $\mathcal{D}$ by depth $a + 1$.

Let $\tau$ be the threshold value of TA at depth $a + 1$. Thus, if $\underline{x}_i$ is the grade of the $(a + 1)$th highest object in list $i$, then $\tau = t(\underline{x}_1, \ldots, \underline{x}_m)$. Let us call an object $R$ *big* if $t(R) \geq \tau$, and otherwise call object $R$ *small*. (Note that these definitions of "big" and "small" are different from those in the proof of Theorem 6.1.)

We now show that every member $R$ of $Y$ is big. Let $x'_i$ be some grade in the top $a + 1$ grades in list $i$ that is not the grade in list $i$ of any object seen by $\mathcal{A}$. There is such a grade, since all grades in list $i$ are distinct, and $\mathcal{A}$ sees at most $a$ objects. Let $\mathcal{D}'$ agree with $\mathcal{D}$ on all objects seen by A, and let object $V$ have grade $x'_i$ in the $i$th list of $\mathcal{D}'$, for $1 \leq i \leq m$. Hence, the grade of $V$ in $\mathcal{D}'$ is $t(x'_1, \ldots, x'_m) \geq \tau$. Since $V$ was unseen, and since $V$ is assigned grades in each list in $\mathcal{D}'$ below the level that $\mathcal{A}$ reached by sorted access, it follows that algorithm $\mathcal{A}$ performs exactly the same, and in particular gives the same output, for databases $\mathcal{D}$ and $\mathcal{D}'$. Therefore, algorithm $\mathcal{A}$ has $R$, but not $V$, in its output for database $\mathcal{D}'$. By correctness of $\mathcal{A}$, it follows that $R$ is big, as desired.

We claim that every member $R$ of $Y$ is one of the top $a + 1$ members of some list $i$ (and so is seen by TA by depth $a + 1$). Assume by way of contradiction that $R$ is not one of the top $a + 1$ members of list $i$, for $1 \leq i \leq m$. By our assumptions that the aggregation function $t$ is strictly monotone. and that $\mathcal{D}$ satisfies the distinctness property, it follows easily that $R$ is small. We already showed that every member of $Y$ is big. This contradiction proves the claim. It follows that TA halts by depth $a + 1$, as desired. □

In the proof of Theorem 6.5, we showed that under the assumptions of Theorem 6.5 (strict monotonicity and the distinctness property) the optimality ratio of TA is at most $cm^2$, where $c = \max\{c_R/c_S, c_S/c_R\}$.

---

[12]This should not be confused with the aggregation function being both strict and monotone. We apologize for the clash in terminology, which exists for historical reasons.

In Theorem 9.2, we give an aggregation function that is strictly monotone such that no deterministic algorithm can have an optimality ratio of less than $\frac{m-2}{2}\frac{c_R}{c_S}$. So in our case of greatest interest, where $c_R \geq c_S$, there is a gap of around a factor of $2m$ in the upper and lower bounds.

The proofs of Theorems 6.1 and 6.5 have several nice properties:

- The proofs would still go through if we were in a scenario where, whenever a random access of object $R$ in list $i$ takes place, we learn not only the grade of $R$ in list $i$, but also the relative rank. Thus, TA is instance optimal even when we allow **A** to include also algorithms that learn and make use of such relative rank information.

- As we shall see, we can prove the instance optimality among approximation algorithms of an approximation version of TA, under the assumptions of Theorem 6.1, with only a small change to the proof (as we shall see, such a theorem does not hold under the assumptions of Theorem 6.5).

## 6.1 Treating k and m as Constants

In Theorems 6.1 and 6.5 about the instance optimality of TA, we are treating $k$ (where we are trying to find the top $k$ answers) and $m$ (the number of sorted lists) as constants. We now discuss these assumptions.

We begin first with the assumption that $k$ is constant. As in the proofs of Theorems 6.1 and 6.5, let $a$ be the number of accesses by an algorithm $\mathcal{A} \in \mathbf{A}$. If $a \geq k$, then there is no need to treat $k$ as a constant. Thus, if we were to restrict the class **A** of algorithms to contain only algorithms that make at least $k$ accesses to find the top $k$ answers, then there would be no need to assume that $k$ is constant. How can it arise that an algorithm $\mathcal{A}$ can find the top $k$ answers without making at least $k$ accesses, and in particular without accessing at least $k$ objects? It must then happen that either there are at most $k$ objects in the database, or else every object $R$ that $\mathcal{A}$ has not seen has the same overall grade $t(R)$. The latter will occur, for example, if $t$ is a constant function. Even under these circumstances, it is still not reasonable in some contexts (such as certain database contexts) to allow an algorithm $\mathcal{A}$ to output an object as a member of the top $k$ objects without ever having seen it: how would the algorithm even know the name of the object? This is similar to an issue we raised earlier about wild guesses.

What about the assumption that $m$ is constant? As we noted earlier, this is certainly a reasonable assumption, since $m$ is the number of arguments of the aggregation function, which we are of course taking to be fixed. In the case of the assumptions of Theorem 6.1 (no wild guesses), Corollary 6.2 tells us that at least for strict aggregation functions, this dependence on $m$ is inevitable. Similarly, in the case of the assumptions of Theorem 6.5 (strict monotonicity and the distinctness property), Theorem 9.2 tells us that at least for certain aggregation functions, this dependence on $m$ is inevitable.

## 6.2 Turning TA into an Approximation Algorithm, and Allowing Early Stopping

TA can easily be modified to be an *approximation algorithm*. It can then be used in situations where we care only about the *approximately* top $k$ answers. Thus, let $\theta > 1$ be given. Define a *$\theta$-approximation to the top $k$ answers (for $t$ over database $\mathcal{D}$)* to be a collection of $k$ objects (and their grades) such that for each $y$ among these $k$ objects and each $z$ not among these $k$ objects, $\theta t(y) \geq t(z)$. We can modify TA to find a $\theta$-approximation to the top $k$ answers by modifying the stopping rule in Step 2 to say "As soon as at least $k$ objects have been seen whose grade is at least equal to $\tau/\theta$, then halt." Let us call this approximation algorithm TA$_\theta$.

| $L_1$ | $L_2$ |
|---|---|
| $(1, \cdot)$ | $(2n+1, \cdot)$ |
| $(2, \cdot)$ | $(2n, \cdot)$ |
| ... | ... |
| $(n+1, \frac{1}{\theta})$ | $(n+1, \frac{1}{\theta})$ |
| $(n+2, \frac{1}{2\theta^2})$ | $(n, \frac{1}{2\theta^2})$ |
| $(n+3, \cdot)$ | $(n-1, \cdot)$ |
| ... | ... |
| $(2n+1, \cdot)$ | $(1, \cdot)$ |

Figure 2: Database for Example 6.8

**Theorem 6.6:** *Assume that $\theta > 1$ and that the aggregation function $t$ is monotone. Then TA$_\theta$ correctly finds a $\theta$-approximation to the top $k$ answers for $t$.*

**Proof:** This follows from a straightforward modification of the proof of Theorem 4.1. $\square$

The next theorem says that when we restrict attention to algorithms that do not make wild guesses, then TA$_\theta$ is instance optimal.

**Theorem 6.7:** *Assume that $\theta > 1$ and that the aggregation function $t$ is monotone. Let **D** be the class of all databases. Let **A** be the class of all algorithms that find a $\theta$-approximation to the top $k$ answers for $t$ for every database and that do not make wild guesses. Then TA$_\theta$ is instance optimal over **A** and **D**.*

**Proof:** The proof of Theorem 6.1 carries over verbatim provided we modify the definition of an object $R$ being "big" to be that $\theta t(R) \geq \tau_\mathcal{A}$. $\square$

Theorem 6.7 shows that the analog of Theorem 6.1 holds for TA$_\theta$. The next example, which is a modification of Example 6.3, shows that the analog of Theorem 6.5 does *not* hold for TA$_\theta$. One interpretation of these results is that Theorem 6.1 is sufficiently robust that it can survive the perturbation of allowing approximations, whereas Theorem 6.5 is not.

**Example 6.8:** Assume that $\theta > 1$, that there are $2n+1$ objects, which we will call simply $1, 2, \ldots, 2n+1$, and that there are two lists $L_1$ and $L_2$ (see Figure 2).[13] Assume that in list $L_1$, the grades are assigned so that all grades are different, the ordering of the objects by grade is $1, 2, \ldots, 2n+1$, object $n+1$ has the grade $1/\theta$, and object $n+2$ has the grade $1/(2\theta^2)$. Assume that in list $L_2$, the grades are assigned so that all grades are different, the ordering of the objects by grade is $2n+1, 2n, \ldots, 1$ (the reverse of the ordering in $L_1$), object $n+1$ has the grade $1/\theta$, and object $n$ has the grade $1/(2\theta^2)$. Assume that the aggregation function is min, and that $k = 1$ (so that we are interested in finding a $\theta$-approximation to the top answer). The (overall) grade of each object other than object $n+1$ is at most $\alpha = 1/(2\theta^2)$. Since $\theta\alpha = 1/(2\theta)$, which is less than the grade $1/\theta$ of object $n+1$, it follows that the unique object that can be returned by an algorithm such as TA$_\theta$ that correctly finds a $\theta$-approximation to the top answer is the object $n + 1$.

---
[13]In this and later figures, each centered dot represents a value that it is not important to give explicitly.

An algorithm that makes a wild guess and asks for the grade of object $n + 1$ in both lists would determine the correct answer and be able to halt safely after two random accesses and no sorted accesses. The algorithm could halt safely, since it "knows" that it has found an object $R$ such that $\theta t(R) = 1$, and so $\theta t(R)$ is at least as big as every possible grade. However, under sorted access for list $L_1$, the algorithm TA$_\theta$ would see the objects in the order $1, 2, \ldots, 2n + 1$, and under sorted access for list $L_2$, the algorithm TA$_\theta$ would see the objects in the reverse order. Since the winning object $n + 1$ is in the middle of both sorted lists, it follows that at least $n + 1$ sorted accesses would be required before TA$_\theta$ would even see the winning object. □

Just as we converted Example 6.3 into Theorem 6.4, we can convert Example 6.8 into the following theorem.

**Theorem 6.9:** *Assume that $\theta > 1$. Let **D** be the class of all databases that satisfy the distinctness property. Let **A** be the class of all algorithms that find a $\theta$-approximation to the top answer for min for every database in **D**. There is no deterministic algorithm (or even probabilistic algorithm that never makes a mistake) that is instance optimal over **A** and **D**.*

**Early stopping of TA:** It is straightforward to modify TA$_\theta$ into an interactive process where at all times the system can show the user the current top $k$ list along with a guarantee about the degree of approximation to the correct answer. At any time, the user can decide, based on this guarantee, whether he would like to stop the process. Thus, let $\beta$ be the grade of the $k$th (bottom) object in the current top $k$ list, let $\tau$ be the current threshold value, and let $\theta = \tau/\beta$. If the algorithm is stopped early, we have $\theta > 1$. It is easy to see that similarly to the situation of Theorem 6.6, the current top $k$ list is then a $\theta$-approximation to the top $k$ answers. Thus, the user can be shown the current top $k$ list and the number $\theta$, with a guarantee that he is being shown a $\theta$-approximation to the top $k$ answers.

# 7 Restricting Sorted Access

Bruno, Gravano, and Marian [**?**] discuss a scenario where it is not possible to access certain of the lists under sorted access. They give a nice example where the user wants to get information about restaurants. The user has an aggregation function that gives a score to each restaurant based on how good it is, how inexpensive it is, and how close it is. In this example, the Zagat-Review web site gives ratings of restaurants, the NYT-Review web site gives prices, and the MapQuest web site gives distances. Only the Zagat-Review web site can be accessed under sorted access (with the best restaurants at the top of the list).

Let $Z$ be the set of indices $i$ of those lists $L_i$ that can be accessed under sorted access. We assume that $Z$ is nonempty, that is, that at least one of the lists can be accessed under sorted access. We take $m'$ to be the cardinality $|Z|$ of $Z$ (and as before, take $m$ to be the total number of sorted lists). Define TA$_Z$ to be the following natural modification of TA, that deals with the restriction on sorted access.

1. Do sorted access in parallel to each of the $m'$ sorted lists $L_i$ with $i \in Z$. As an object $R$ is seen under sorted access in some list, do random access as needed to the other lists to find the grade $x_i$ of object $R$ in every list $L_i$. Then compute the grade $t(R) = t(x_1, \ldots, x_m)$ of object $R$. If this grade is one of the $k$ highest we have seen, then remember object $R$ and its grade $t(R)$ (ties are broken arbitrarily, so that only $k$ objects and their grades need to be remembered at any time).

2. For each list $L_i$ with $i \in Z$, let $\underline{x}_i$ be the grade of the last object seen under sorted access. For each list $L_i$ with $i \notin Z$, let $\underline{x}_i = 1$. Define the *threshold value* $\tau$ to be $t(\underline{x}_1, \ldots, \underline{x}_m)$. As soon as at least $k$ objects have been seen whose grade is at least equal to $\tau$, then halt.[14]

3. Let $Y$ be a set containing the $k$ objects that have been seen with the highest grades. The output is then the graded set $\{(R, t(R)) \mid R \in Y\}$.

In the case where $|Z| = 1$, algorithm TA$_Z$ is essentially the same as the algorithm TA-Adapt in [?].

In footnote 5, we noted that each of the algorithms in this paper where there is "sorted access in parallel" remain correct even when sorted access is not in lockstep. Algorithm TA$_Z$ provides an extreme example, where only some of the sorted lists are accessed under sorted access, and the remaining sorted lists are accessed under random access only.

We now show that Theorem 6.1, which says that TA is instance optimal when we restrict attention to algorithms that do not make wild guesses, and Corollary 6.2, which says that the optimality ratio of TA is best possible when we restrict attention to algorithms that do not make wild guesses, both generalize to hold for TA$_Z$. What about our other theorem about instance optimality of TA (Theorem 6.5), which says that TA is instance optimal when the aggregation function $t$ is strictly monotone and the class of legal databases satisfies the distinctness property? Interestingly enough, we shall show (Example 7.3) that this latter theorem does not generalize to TA$_Z$.

**Theorem 7.1:** *Assume that the aggregation function $t$ is monotone. Let **D** be the class of all databases. Let **A** be the class of all algorithms that correctly find the top $k$ answers for $t$ for every database and that do not make wild guesses, where the only lists that may be accessed under sorted access are those lists $L_i$ with $i \in Z$. Then TA$_Z$ is instance optimal over **A** and **D**.*

**Proof:** The proof is essentially the same as the proof of Theorem 6.1, except for the bookkeeping. Assume that $\mathcal{A} \in \mathbf{A}$, and that algorithm $\mathcal{A}$ is run over database $\mathcal{D}$. Assume that algorithm $\mathcal{A}$ halts at depth $d$ (that is, if $d_i$ is the number of objects seen under sorted access to list $i$, for $1 \leq i \leq m$, then $d = \max_i d_i$). Assume that $\mathcal{A}$ sees $a$ distinct objects (some possibly multiple times). Since $\mathcal{A}$ makes no wild guesses, and sees $a$ distinct objects, it must make at least $a$ sorted accesses, and so its middleware cost is at least $ac_S$. By the same proof as that of Theorem 6.1, it follows that TA$_Z$ halts on $\mathcal{D}$ by depth $a + k$. Hence, the middleware cost of TA$_Z$ is at most $(a+k)m'c_S + (a+k)m'(m-1)c_R$, which is $am'c_S + am'(m-1)c_R$ plus an additive constant of $km'c_S + km'(m-1)c_R$. So the optimality ratio of TA$_Z$ is at most $\frac{am'c_S + am'(m-1)c_R}{ac_S} = m' + m'(m-1)c_R/c_S$. $\square$

The next result, which is analogous to Corollary 6.2, is a corollary of the proof of Theorem 7.1 and of a lower bound in Section 9.

---

[14] As we shall see in Example 7.3, even though there are at least $k$ objects, it is possible that after seeing the grade of every object in every list, and thus having done sorted access to every object in every list $L_i$ with $i \in Z$, there are not at least $k$ objects with a grade that is at least equal to the final threshold value $\tau$. In this situation, we say that TA$_Z$ halts after it has seen the grade of every object in every list. This situation cannot happen with TA.

| $L_1$ | $L_2$ | $L_3$ |
|---|---|---|
| $(R, 1)$ | ... | $(R, 1)$ |
| ... | $(R, 0.6)$ | ... |
| $(\cdot, 0.7)$ | ... | ... |

Figure 3: Database for Example 7.3

**Corollary 7.2:** *Let $t$ be an arbitrary monotone, strict aggregation function with $m$ arguments. Assume that $|Z| = m'$. Let $\mathbf{D}$ be the class of all databases. Let $\mathbf{A}$ be the class of all algorithms that correctly find the top $k$ answers for $t$ for every database, and that do not make wild guesses, where the only lists that may be accessed under sorted access are those lists $L_i$ with $i \in Z$. Then $TA_Z$ is instance optimal over $\mathbf{A}$ and $\mathbf{D}$, with optimality ratio $m' + m'(m-1)c_R/c_S$. No deterministic algorithm has a lower optimality ratio.*

**Proof:** In the proof of Theorem 7.1, it is shown that $TA_Z$ has an optimality ratio of at most $m' + m'(m-1)c_R/c_S$ for an arbitrary monotone aggregation function, The lower bound follows from a minor variation of the proof of Theorem 9.1, where we take $\psi = (dm'-1)c_S + (dm'-1)(m-1)c_R$. The simple details are left to the reader. $\square$

Theorem 6.5 says that if the aggregation function $t$ is strictly monotone, and if the class of legal databases satisfies the distinctness property, then TA is instance optimal. We now show by example that the analogous result fails for $TA_Z$. In fact, we shall show that $TA_Z$ need not be instance optimal even if we assume not only that aggregation function $t$ is strictly monotone, and that the class of legal databases satisfies the distinctness property, but in addition we assume that the aggregation function $t$ is strict, and that no wild guesses are allowed.

**Example 7.3:** Assume that the database satisfies the distinctness property, that there are only three sorted lists $L_1$, $L_2$, and $L_3$ (see Figure 3), and that $Z = \{1\}$ (so that only $L_1$ may be accessed under sorted access). Let $t$ be the aggregation function where $t(x, y, z) = \min\{x, y\}$ if $z = 1$, and $t(x, y, z) = (\min\{x, y, z\})/2$ if $z \neq 1$. It is easy to see that $t$ is strictly monotone and strict. Assume that we are interested in finding the top answer (i.e., $k = 1$).

Assume that object $R$ has grade 1 in lists $L_1$ and $L_3$, and grade 0.6 in list $L_2$. Hence $t(R) = 0.6$. Note that for each object $R'$ other than $R$, necessarily the grade of $R'$ in $L_3$ is not 1 (by the distinctness property), and so $t(R') \leq 0.5$. Therefore, $R$ is the unique top object.

Assume that the minimum grade in list $L_1$ is 0.7. It follows that the threshold value is never less than 0.7. Therefore, $TA_Z$ does not halt until it has seen the grade of every object in every list. However, let $\mathcal{A}$ be an algorithm that does sorted access to the top object $R$ in list $L_1$ and random access to $R$ in lists $L_2$ and $L_3$, and then halts and announces that $R$ is the top object. Algorithm $\mathcal{A}$ does only one sorted access and two random accesses on this database. It is safe for algorithm $\mathcal{A}$ to halt, since it "knows" that object $R$ has grade 0.6 and that no other object can have grade bigger than 0.5. Since there can be an arbitrarily large number of objects, it follows that $TA_Z$ is not instance optimal. Hence, the analogue of Theorem 6.5 fails for $TA_Z$.

It is instructive to understand "what goes wrong" in this example and why this same problem does not also cause Theorems 6.5 or 7.1 to fail. Intuitively, what goes wrong in this example is that the threshold value is too conservative an estimate as an upper bound on the grade of unseen objects. By contrast, in the case of Theorem 7.1, some unseen object may have an overall grade equal to the threshold value, so the threshold value is not too conservative an estimate. In the case of Theorem 6.5, an analysis of the proof shows that we consider the threshold value at depth $a+1$ rather than depth $a$. Intuitively, although the threshold value may be too conservative an estimate, the threshold value one extra level down is not. □

## 8 Minimizing Random Access

Thus far in this paper, we have not been especially concerned about the number of random accesses. For every sorted access in TA, up to $m-1$ random accesses take place. Recall that if $s$ is the number of sorted accesses, and $r$ is the number of random accesses, then the middleware cost is $sc_S + rc_R$, for some positive constants $c_S$ and $c_R$. Our notion of instance optimality ignores constant factors like $m$ and $c_R$ (they are simply multiplicative factors in the optimality ratio). Hence, there has been no motivation so far to concern ourself with the number of random accesses.

There are, however, some scenarios where we must pay attention to the number of random accesses. The first scenario is where random accesses are impossible (which corresponds to $c_R = \infty$). As we discussed in Section 2, an example of this first scenario arises when the middleware system is a text retrieval system, and the sorted lists correspond to the results of search engines. Another scenario is where random accesses are not impossible, but simply expensive relative to sorted access. An example of this second scenario arises when the costs correspond to disk access (sequential versus random). Then we would like the optimality ratio to be independent of $c_R/c_S$. That is, if we allow $c_R$ and $c_S$ to vary, instead of treating them as constants, we would still like the optimality ratio to be bounded.

In this section we describe algorithms that do not use random access frivolously. We give two algorithms. One uses no random accesses at all, and hence is called NRA ("No Random Access"). The second algorithm takes into account the cost of a random access. It is a combination of NRA and TA, and so we call it CA ("Combined Algorithm").

Both algorithms access the information in a natural way, and, in the spirit of the knowledge-based programs of Section 4, halt when they know that no improvement can take place. In general, at each point in an execution of these algorithms where a number of sorted and random accesses have taken place, for each object $R$ there is a subset $S(R) = \{i_1, i_2, \ldots, i_\ell\} \subseteq \{1, \ldots, m\}$ of the fields of $R$ where the algorithm has determined the values $x_{i_1}, x_{i_2}, \ldots, x_{i_\ell}$ of these fields. Given this information, we define functions of this information that are lower and upper bounds on the value $t(R)$ can obtain. The algorithm proceeds until there are no more candidates whose current upper bound is better than the current $k$th largest lower bound.

**Lower Bound:** Given an object $R$ and subset $S(R) = \{i_1, i_2, \ldots, i_\ell\} \subseteq \{1, \ldots, m\}$ of known fields of $R$, with values $x_{i_1}, x_{i_2}, \ldots, x_{i_\ell}$ for these known fields, we define $W_S(R)$ (or W(R) if the subset $S = S(R)$ is understood from the context) as the minimum (or *worst*) value the aggregation function $t$ can attain for object $R$. When $t$ is monotone, this minimum value is obtained by substituting for each missing field $i \in \{1, \ldots, m\} \setminus S$ the value 0, and applying $t$ to the result. For example, if $S = \{1, \ldots, \ell\}$, then $W_S(R) = t(x_1, x_2, \ldots, x_\ell, 0, \ldots, 0)$. The following property is immediate from the definition:

**Proposition 8.1:** *If $S$ is the set of known fields of object $R$, then $t(R) \geq W_S(R)$.*

In other words, $W(R)$ represents a lower bound on $t(R)$. Is it the best possible? Yes, unless we have additional information, such as that the value 0 does not appear in the lists. In general, as an algorithm progresses and we learn more fields of an object $R$, its $W$ value becomes larger (or at least not smaller). For some aggregation functions $t$ the value $W(R)$ yields no knowledge until $S$ includes all fields: for instance if $t$ is min, then $W(R)$ is 0 until all values are discovered. For other functions it is more meaningful. For instance, when $t$ is the median of three fields, then as soon as two of them are known $W(R)$ is at least the smaller of the two.

**Upper Bound:** The best value an object can attain depends on other information we have. We will use only the *bottom values* in each field, defined as in TA: $\underline{x}_i$ is the last (smallest) value obtained via sorted access in list $L_i$. Given an object $R$ and subset $S(R) = \{i_1, i_2, \ldots, i_\ell\} \subseteq \{1, \ldots, m\}$ of known fields of $R$, with values $x_{i_1}, x_{i_2}, \ldots, x_{i_\ell}$ for these known fields, we define $B_S(R)$ (or B(R) if the subset $S$ is understood from the context) as the maximum (or *best*) value the aggregation function $t$ can attain for object $R$. When $t$ is monotone, this maximum value is obtained by substituting for each missing field $i \in \{1, \ldots, m\} \backslash S$ the value $\underline{x}_i$, and applying $t$ to the result. For example, if $S = \{1, \ldots, \ell\}$, then $B_S(R) = t(x_1, x_2, \ldots, x_\ell, \underline{x}_{\ell+1}, \ldots, \underline{x}_m)$. The following property is immediate from the definition:

**Proposition 8.2:** *If $S$ is the set of known fields of object $R$, then $t(R) \leq B_S(R)$.*

In other words, $B(R)$ represents an upper bound on the value $t(R)$ (or the *best* value $t(R)$ can be), given the information we have so far. Is it the best upper bound? If the lists may each contain equal values (which in general we assume they can), then given the information we have it is possible that $t(R) = B_S(R)$. If the distinctness property holds (equalities are not allowed in a list), then for continuous aggregation functions $t$ it is the case that $B(R)$ is the best upper bound on the value $t$ can have on $R$. In general, as an algorithm progresses and we learn more fields of an object $R$ and the bottom values $\underline{x}_i$ decrease, $B(R)$ can only decrease (or remain the same).

An important special case is an object $R$ that has not been encountered at all. In this case $B(R) = t(\underline{x}_1, \underline{x}_2, \ldots, \underline{x}_m)$. Note that this is the same as the threshold value in TA.

## 8.1 No Random Access Algorithm—NRA

As we have discussed, there are situations where random accesses are impossible. We now consider algorithms that make no random accesses. Since random accesses are impossible, in this section we change our criterion for the desired output. In earlier sections, we demanded that the output be the "top $k$ answers", which consists of the top $k$ objects, along with their (overall) grades. In this section, we make the weaker requirement that the output consist of the top $k$ objects, without their grades. The reason is that, since random access is impossible, it may be much cheaper (that is, require many fewer accesses) to find the top $k$ objects without their grades. This is because, as we now show by example, we can sometimes obtain enough partial information about grades to know that an object is in the top $k$ objects without knowing its exact grade.

**Example 8.3:** Consider the following scenario, where the aggregation function is the average, and where $k = 1$ (so that we are interested only in the top object). There are only two sorted lists $L_1$ and $L_2$ (see Figure 4), and the grade of every object in both $L_1$ and $L_2$ is 1/3, except that object $R$ has grade 1 in $L_1$ and grade 0 in $L_2$. After two sorted accesses to $L_1$ and one sorted access to $L_2$, there is enough information to know that object $R$ is the top object (its average grade is at least 1/2, and every other object has average grade at most 1/3). If we wished to find the grade of object $R$, we would need to do sorted access to all of $L_2$. □

| $L_1$ |
|---|
| $(R, 1)$ |
| $(\cdot, \frac{1}{3})$ |
| $\ldots$ |
| $(\cdot, \frac{1}{3})$ |

| $L_2$ |
|---|
| $(\cdot, \frac{1}{3})$ |
| $\ldots$ |
| $(\cdot, \frac{1}{3})$ |
| $(R, 0)$ |

Figure 4: Database for Example 8.3

Note that we are requiring only that the output consist of the top $k$ objects, with no information being given about the sorted order (sorted by grade). If we wish to know the sorted order, this can easily be determined by finding the top object, the top 2 objects, etc. Let $C_i$ be the cost of finding the top $i$ objects. It is interesting to note that there is no necessary relationship between $C_i$ and $C_j$ for $i < j$. For example, in Example 8.3, we have $C_1 < C_2$. If we were to modify Example 8.3 so that there are two objects $R$ and $R'$ with grade 1 in $L_1$, where the grade of $R$ in $L_2$ is 0, and the grade of $R'$ in $L_2$ is 1/4 (and so that, as before, all remaining grades of all objects in both lists is 1/3), then $C_2 < C_1$.

The cost of finding the top $k$ objects in sorted order is at most $k \max_i C_i$. Since we are treating $k$ as a constant, it follows easily that we can convert our instance optimal algorithm (which we shall give shortly) for finding the top $k$ objects into an instance optimal algorithm for finding the top $k$ objects in sorted order. In practice, it is usually good enough to know the top $k$ objects in sorted order, without knowing the grades. In fact, the major search engines on the web no longer give grades (possibly to prevent reverse engineering).

The algorithm NRA is as follows.

1. Do sorted access in parallel to each of the $m$ sorted lists $L_i$. At each depth $d$ (when $d$ objects have been accessed under sorted access in each list):

    - Maintain the bottom values $\underline{x}_1^{(d)}, \underline{x}_2^{(d)}, \ldots, \underline{x}_m^{(d)}$ encountered in the lists.
    - For every object $R$ with discovered fields $S = S^{(d)}(R) \subseteq \{1, \ldots, m\}$, compute the values $W^{(d)}(R) = W_S(R)$ and $B^{(d)}(R) = B_S(R)$. (For objects $R$ that have not been seen, these values are virtually computed as $W^{(d)}(R) = t(0, \ldots, 0)$, and $B^{(d)}(R) = t(\underline{x}_1, \underline{x}_2, \ldots, \underline{x}_m)$, which is the threshold value.)
    - Let $T_k^{(d)}$, the current top $k$ list, contain the $k$ objects with the largest $W^{(d)}$ values seen so far (and their grades); if two objects have the same $W^{(d)}$ value, then ties are broken using the $B^{(d)}$ values, such that the object with the highest $B^{(d)}$ value wins (and arbitrarily among objects that tie for the highest $B^{(d)}$ value). Let $M_k^{(d)}$ be the $k$th largest $W^{(d)}$ value in $T_k^{(d)}$.

2. Call an object $R$ *viable* if $B^{(d)}(R) > M_k^{(d)}$. Halt when (a) at least $k$ distinct objects have been seen (so that in particular $T_k^{(d)}$ contains $k$ objects) and (b) there are no viable objects left outside $T_k^{(d)}$, that is, when $B^{(d)}(R) \leq M_k^{(d)}$ for all $R \notin T_k^{(d)}$. Return the objects in $T_k^{(d)}$.

We now show that NRA is correct for each monotone aggregation function $t$.

**Theorem 8.4:** *If the aggregation function $t$ is monotone, then NRA correctly finds the top $k$ objects.*

**Proof:** Assume that NRA halts after $d$ sorted accesses to each list, and that $T_k^{(d)} = \{R_1, R_2, \ldots, R_k\}$. Thus, the objects output by NRA are $R_1, R_2, \ldots, R_k$. Let $R$ be an object not among $R_1, R_2, \ldots, R_k$. We must show that $t(R) \leq t(R_i)$ for each $i$.

Since the algorithm halts at depth $d$, we know that $R$ is nonviable at depth $d$, that is, $B^{(d)}(R) \leq M_k^{(d)}$. Now $t(R) \leq B^{(d)}(R)$ (Proposition 8.2). Also for each of the $k$ objects $R_i$ we have $M_k^{(d)} \leq W^{(d)}(R_i) \leq t(R_i)$ (from Proposition 8.1 and the definition of $M_k^{(d)}$). Combining the inequalities we have shown, we have
$$t(R) \leq B^{(d)}(R) \leq M_k^{(d)} \leq W^{(d)}(R_i) \leq t(R_i)$$
for each $i$, as desired. $\square$

Note that the tie-breaking mechanism was not needed for correctness (but will be used for instance optimality). We now show instance optimality of NRA over all algorithms that do not use random access:

**Theorem 8.5:** *Assume that the aggregation function $t$ is monotone. Let $\mathbf{D}$ be the class of all databases. Let $\mathbf{A}$ be the class of all algorithms that correctly find the top $k$ objects for $t$ for every database and that do not make random accesses. Then NRA is instance optimal over $\mathbf{A}$ and $\mathbf{D}$.*

**Proof:** Assume $\mathcal{A} \in \mathbf{A}$. If algorithm NRA halts at depth $d$, and NRA saw at least $k$ distinct objects for the first time by depth $d$, then NRA makes only a constant number of accesses (at most $km^2$) on that database. So suppose that on some database $\mathcal{D}$, algorithm NRA halts at depth $d$, and that NRA saw at least $k$ distinct objects by depth $d-1$. We claim that $\mathcal{A}$ must get to depth $d$ in *at least one of the lists*. It then follows that the optimality ratio of NRA is at most $m$, and the theorem follows. Suppose the claim fails; then from the fact that algorithm NRA did not halt at depth $d-1$ there is an object $R \notin T_k^{(d-1)}$ such that $B^{(d-1)}(R) > M_k^{(d-1)}$. We know that $W^{(d-1)}(R) \leq M_k^{(d-1)}$, since $R \notin T_k^{(d-1)}$ Further, we know from the tie-breaking mechanism that if $W^{(d-1)}(R) = M_k^{(d-1)}$, then for each $R_i \in T_k^{(d-1)}$ such that $W^{(d)}(R_i) = M_k^{(d)}$ necessarily $B^{(d-1)}(R_i) \geq B^{(d-1)}(R)$.

There are now two cases, depending on whether or not algorithm $\mathcal{A}$ outputs $R$ as one of the top $k$ objects. In either case, we construct a database on which $\mathcal{A}$ errs.

*Case 1:* Algorithm $\mathcal{A}$ outputs $R$ as one of the top $k$ objects. We construct a database $\mathcal{D}'$ where $\mathcal{A}$ errs as follows. Database $\mathcal{D}'$ is identical to $\mathcal{D}$ up to depth $d-1$ (that is, for each $i$ the top $d-1$ objects and their grades are the same in list $L_i$ for $\mathcal{D}'$ as for $\mathcal{D}$). For each $R_i$ and for each missing field $j \in \{1, \ldots, m\} \setminus S^{(d-1)}(R_i)$ assign value $\underline{x}_j^{(d-1)}$. For the object $R$ assign all of the missing fields in $\{1, \ldots, m\} \setminus S^{(d-1)}(R)$ the value 0. We now show that $t(R) < t(R_j)$ for each $j$ with $1 \leq j \leq k$. Hence, $R$ is *not* one of the top $k$ objects, and so algorithm $\mathcal{A}$ erred. First, we have

$$t(R) = W^{(d-1)}(R) \leq M_k^{(d-1)} \tag{3}$$

Also, for all $i$ with $1 \leq i \leq k$ we have

$$M_k^{(d-1)} \leq W^{(d-1)}(R_i) \leq B^{(d-1)}(R_i) = t(R_i). \tag{4}$$

If $W^{(d-1)}(R) < M_k^{(d-1)}$, then we have from (3) and (4) that $t(R) < t(R_i)$ for each $i$, as desired. So assume that $W^{(d-1)}(R) = M_k^{(d-1)}$. Again, we wish to show that $t(R) < t(R_i)$ for each $i$. We consider separately in two subcases those $i$ where $M_k^{(d-1)} = W^{(d-1)}(R_i)$ and those where $M_k^{(d-1)} \neq W^{(d-1)}(R_i)$.

*Subcase 1:* $M_k^{(d-1)} = W^{(d-1)}(R_i)$. Then $t(R) \leq M_k^{(d-1)} < B^{(d-1)}(R) \leq B^{(d-1)}(R_i) = t(R_i)$, as desired, where the last inequality follows from the tie-breaking mechanism.

*Subcase 2:* $M_k^{(d-1)} \neq W^{(d-1)}(R_i)$, and so $M_k^{(d-1)} < W^{(d-1)}(R_i)$. From the inequalities in (4), we see that $M_k^{(d-1)} < t(R_i)$. So by (3), we have $t(R) < t(R_i)$, as desired.

*Case 2:* Algorithm $\mathcal{A}$ does not output $R$ as one of the top $k$ objects. We construct a database $\mathcal{D}''$ where $\mathcal{A}$ errs as follows. Database $\mathcal{D}''$ is identical to $\mathcal{D}$ up to depth $d-1$. At depth $d$ it gives each missing field $i \in \{1, \ldots, m\} \setminus S^{(d-1)}(R)$ of $R$ the value $\underline{x}_i^{(d-1)}$. For all remaining missing fields, including missing fields of $R_1, \ldots, R_k$, assign the value 0. Now $t(R) = B^{(d-1)}(R) > M_k^{(d-1)}$, whereas (a) for at least one $R_i$ (namely, that $R_i$ where $W^{(d)}(R_i) = M_k^{(d)}$) we have $t(R_i) = M_k^{(d-1)}$, and (b) for each object $R'$ not among $R_1, R_2, \ldots, R_k$ or $R$ we have that $t(R') \leq M_k^{(d-1)}$. Hence, algorithm $\mathcal{A}$ erred in not outputting $R$ as one of the top $k$ objects. □

Note that the issue of "wild guesses" is not relevant here, since we are restricting our attention to algorithms that make no random accesses (and hence no wild guesses).

The next result, which is analogous to Corollaries 6.2 and 7.2 is a corollary of the proof of Theorem 8.4 and of a lower bound in Section 9. Specifically, in the proof of Theorem 8.4, we showed that the optimality ratio of NRA is at most $m$. The next result says that if the aggregation function is strict, then the optimality ratio is precisely $m$, and this is best possible.

**Corollary 8.6:** *Let $t$ be an arbitrary monotone, strict aggregation function with $m$ arguments. Let $\mathbf{D}$ be the class of all databases. Let $\mathbf{A}$ be the class of all algorithms that correctly find the top $k$ objects for $t$ for every database and that do not make random accesses. Then NRA is instance optimal over $\mathbf{A}$ and $\mathbf{D}$, with optimality ratio $m$. No deterministic algorithm has a lower optimality ratio.*

**Proof:** In the proof of Theorem 8.4, it is shown that NRA has an optimality ratio of at most $m$ for an arbitrary monotone aggregation function, The lower bound follows from Theorem 9.5. □

**Remark 8.7:** Unfortunately, the execution of NRA may require a lot of bookkeeping at each step, since when NRA does sorted access at depth $\ell$ (for $1 \leq \ell \leq d$), the value of $B^{(\ell)}(R)$ must be updated for every object $R$ seen so far. This may be up to $\ell m$ updates for each depth $\ell$, which yields a total of $\Omega(d^2 m)$ updates by depth $d$. Furthermore, unlike TA, it no longer suffices to have bounded buffers. However, for a specific function like min it is possible that by using appropriate data structures the computation can be greatly simplified. This is an issue for further investigation. □

## 8.2 Taking into Account the Random Access Cost

We now present the combined algorithm CA that does use random accesses, but takes their cost (relative to sorted access) into account. As before, let $c_S$ be the cost of a sorted access and $c_R$ be the cost of a random access. The middleware cost of an algorithm that makes $s$ sorted accesses and $r$ random ones is $sc_S + rc_R$. We know that TA is instance optimal; however, the optimality ratio is a function of the relative cost of a random access to a sorted access, that is $c_R/c_S$. Our goal in this section is to find an algorithm that is instance optimal and where the optimality ratio is independent of $c_R/c_S$. One can view CA as a merge between TA and NRA. Let $h = \lfloor c_R/c_S \rfloor$. We assume in this section that $c_R \geq c_S$, so that $h \geq 1$. The idea of CA is to run NRA, but every $h$ steps to run a random access phase and update the information (the upper and lower bounds $B$ and $W$) accordingly. As in Section 8.1, in this section we require only that the output consist of the top $k$ objects, without their grades. If we wish to obtain the grades, this requires only a constant number of additional random accesses, and so has no effect on instance optimality.

The algorithm CA is as follows.

1. Do sorted access in parallel to each of the $m$ sorted lists $L_i$. At each depth $d$ (when $d$ objects have been accessed under sorted access in each list):

   - Maintain the bottom values $\underline{x}_1^{(d)}, \underline{x}_2^{(d)}, \ldots, \underline{x}_m^{(d)}$ encountered in the lists.
   - For every object $R$ with discovered fields $S = S^{(d)}(R) \subseteq \{1, \ldots, m\}$, compute the values $W^{(d)}(R) = W_S(R)$ and $B^{(d)}(R) = B_S(R)$. (For objects $R$ that have not been seen, these values are virtually computed as $W^{(d)}(R) = t(0, \ldots, 0)$, and $B^{(d)}(R) = t(\underline{x}_1, \underline{x}_2, \ldots, \underline{x}_m)$, which is the threshold value.)
   - Let $T_k^{(d)}$, the current top $k$ list, contain the $k$ objects with the largest $W^{(d)}$ values seen so far (and their grades); if two objects have the same $W^{(d)}$ value, then ties are broken using the $B^{(d)}$ values, such that the object with the highest $B^{(d)}$ value wins (and arbitrarily among objects that tie for the highest $B^{(d)}$ value). Let $M_k^{(d)}$ be the $k$th largest $W^{(d)}$ value in $T_k^{(d)}$.

2. Call an object $R$ *viable* if $B^{(d)}(R) > M_k^{(d)}$. Every $h = \lfloor c_R/c_S \rfloor$ steps (that is, every time the depth of sorted access increases by $h$), do the following: pick the viable object that has been seen for which *not* all fields are known and whose $B^{(d)}$ value is as big as possible (ties are broken arbitrarily). Perform random accesses for all of its (at most $m - 1$) missing fields. If there is no such object, then do not do a random access on this step[15].

3. Halt when (a) at least $k$ distinct objects have been seen (so that in particular $T_k^{(d)}$ contains $k$ objects) and (b) there are no viable objects left outside $T_k^{(d)}$, that is, when $B^{(d)}(R) \leq M_k^{(d)}$ for all $R \notin T_k^{(d)}$. Return the objects in $T_k^{(d)}$.

Note that if $h$ is very large (say larger than the number of objects in the database), then algorithm CA is the same as NRA, since no random access is performed. If $h = 1$, then algorithm CA is similar to TA, but different in intriguing ways. For each step of doing sorted access in parallel, CA performs random accesses for all of the missing fields of some object. Instead of performing random accesses for all of the missing fields of *some* object, TA performs random accesses for all of the missing fields of *every* object seen in sorted access. Later (Section 8.4), we discuss further CA versus TA.

For moderate values of $h$ it is *not* the case that CA is equivalent to the *intermittent algorithm* that executes $h$ steps of NRA and then one step of TA. (That is, the intermittent algorithm does random accesses in the same time order as TA does, but simply delays them, so that it does random accesses every $h$ steps.) We show later (Section 8.4) an example where the intermittent algorithm performs much worse than CA. The difference between the algorithms is that CA picks "wisely" on which objects to perform the random access, namely, according to their $B^{(d)}$ values. Thus, it is not enough to consider the knowledge-based program of Section 4 to design the instance optimal algorithm CA; we need also a principle as to which objects to perform the random access on. This was not an issue in designing TA, since in that context, random accesses increase the cost by only a constant multiple.

---

[15] The reason for this escape clause is so that CA does not make a wild guess. We now give an example where this escape clause may be invoked. Assume that $k = 2$ and $c_R = c_S$. Assume that on the first round of sorted access in parallel, the same object appears in all of the lists. Then on the first opportunity to do a random access, the escape clause must be invoked, since every field is known for the only object that has been seen. In the proof of Theorem 8.9, we show that if the escape clause is invoked after depth $k$ (that is, after there has been at least $k$ rounds of sorted access in parallel), then CA halts immediately after.

Correctness of CA is essentially the same as for NRA, since the same upper and lower bounds are maintained:

**Theorem 8.8**: *If the aggregation function $t$ is monotone, then CA correctly finds the top $k$ objects.*

In the next section, we consider scenarios under which CA is instance optimal, with the optimality ratio independent of $c_R/c_S$.

## 8.3 Instance Optimality of CA

In Section 4, we gave two scenarios under which TA is instance optimal over **A** and **D**. In the first scenario (from Theorem 6.1), (1) the aggregation function $t$ is monotone; (2) **D** is the class of all databases; and (c) **A** is the class of all algorithms that correctly find the top $k$ objects for $t$ for every database and that do not make wild guesses. In the second scenario (from Theorem 6.5), (1) the aggregation function $t$ is strictly monotone; (2) **D** is the class of all databases that satisfy the distinctness property; and (3) **A** is the class of all algorithms that correctly find the top $k$ objects for $t$ for every database in **D**. We might hope that under either of these two scenarios, CA is instance optimal, with optimality ratio independent of $c_R/c_S$. Unfortunately, this hope is false, in both scenarios. In fact, our theorems say that not only does CA fail to fulfill this hope, but so does every algorithm. In other words, neither of these scenarios is enough to guarantee the existence of an algorithm with optimality ratio independent of $c_R/c_S$. In the case of the first scenario, we obtain this negative result from Theorem 9.1. In the case of the second scenario, we obtain this negative result from Theorem 9.2.

However, we shall show that by slightly strengthening the assumption on $t$ in the second scenario, CA becomes instance optimal, with optimality ratio independent of $c_R/c_S$. Let us say that the aggregation function $t$ is *strictly monotone in each argument* if whenever one argument is strictly increased and the remaining arguments are held fixed, then the value of the aggregation function is strictly increased. That is, $t$ is strictly monotone in each argument if $x_i < x'_i$ implies that

$$\begin{aligned} & t(x_1, \ldots, x_{i-1}, x_i, x_{i+1}, \ldots, x_m) \\ < \ & t(x_1, \ldots, x_{i-1}, x'_i, x_{i+1}, \ldots, x_m). \end{aligned}$$

The average (or sum) is strictly monotone in each argument, whereas min is not.

We now show (Theorem 8.9) that in the second scenario above, if we replace "The aggregation function $t$ is strictly monotone" by "The aggregation function $t$ is strictly monotone in each argument", then CA is instance optimal, with optimality ratio independent of $c_R/c_S$. We shall also show (Theorem 8.10) that the same result holds if instead, we simply take $t$ to be min, even though min is not strictly monotone in each argument.

**Theorem 8.9:** *Assume that the aggregation function $t$ is strictly monotone in each argument. Let $\mathbf{D}$ be the class of all databases that satisfy the distinctness property. Let $\mathbf{A}$ be the class of all algorithms that correctly find the top $k$ objects for $t$ for every database in $\mathbf{D}$. Then CA is instance optimal over $\mathbf{A}$ and $\mathbf{D}$, with optimality ratio independent of $c_R/c_S$.*

**Proof:** Assume $\mathcal{D} \in \mathbf{D}$. Assume that when CA runs on $\mathcal{D}$, it halts after doing sorted access to depth $d$. Thus, CA makes $md$ sorted accesses and $r$ random accesses, where $r \leq md/h$. Note that in CA the two components ($mdc_S$ and $rc_R$) of the cost $mdc_S + rc_R$ are roughly equal, and their sum is at most $2mdc_S$. Assume $\mathcal{A} \in \mathbf{A}$, and that $\mathcal{A}$ makes $d'$ sorted accesses and $r'$ random accesses. The cost that $\mathcal{A}$ incurs is therefore $d'c_S + r'c_R$.

Suppose that algorithm $\mathcal{A}$ announces that the objects $R'_1, R'_2, \ldots, R'_k$ are the top $k$. First, we claim that each $R'_i$ appears in the top $d' + r' + 1$ objects of at least one list $L_j$. Suppose not. Then there is an object $R'_i$ output by $\mathcal{A}$ such that in each list there is a vacancy above $R'_i$ that has not been accessed either by sorted or random access. There is a database $\mathcal{D}'$ identical to $\mathcal{D}$ in all locations accessed by $\mathcal{A}$ but with an object $R' \notin \{R'_1, R'_2, \ldots, R'_k\}$ whose values reside in these vacancies. From the distinctness property, for each field the value for $R'$ is strictly larger than that for $R'_i$, and from strict monotonicity of $t$ we have $t(R') > t(R'_i)$, making $R'$ a mandatory member of the output. (Note: we used only strict monotonicity of $t$ rather than the stronger property of being strictly monotone in each variable.) This is a contradiction. Hence, each $R'_i$ appears in the top $d' + r' + 1$ objects of at least one list $L_j$.

Let $S_k = \min\{t(R'_1), t(R'_2), \ldots, t(R'_k)\}$. Define the set $C$ of objects not output by $\mathcal{A}$ whose $B$ value at step $d' + r' + 1$ of CA (that is, after $d' + r' + 1$ parallel sorted accesses) is more than $S_k$, that is,

$$C = \{R \notin \{R'_1, R'_2, \ldots, R'_k\} | B^{(d'+r'+1)}(R) > S_k\}.$$

We claim that for each object $R \in C$, algorithm $\mathcal{A}$ must use a random access (to determine $R$'s value in some list). Suppose not. Then we show a database $\mathcal{D}'$ on which algorithm $\mathcal{A}$ performs the same as on $\mathcal{D}$ but where $t(R) > S_k$. This is a contradiction, since then $R$ would have to be in the output of $\mathcal{A}$. For each field $i$ of $R$ that is not accessed by $\mathcal{A}$, we assign in $\mathcal{D}'$ the highest value from the top $d' + r' + 1$ locations of $L_i$ that had not been accessed by $\mathcal{A}$; such "free" locations exist by the pigeonhole principal, since $\mathcal{A}$ "touched" at most $d' + r'$ objects. Now each field $i$ of $R$ that is accessed by $\mathcal{A}$ is one of the top $d'$ values in $L_i$, since by assumption $R$ was accessed only under sorted access by $\mathcal{A}$. Also, by construction, in $\mathcal{D}'$ each remaining field $i$ of $R$ is one of the top $d' + r' + 1$ values in $L_i$. So in $\mathcal{D}'$, every field $i$ of $R$ is one of the top $d' + r' + 1$ values in $L_i$. Also, by construction, the value of every field $i$ of $R$ is at least as high in $\mathcal{D}'$ as in $\mathcal{D}$. It follows by monotonicity of $t$ that the value of $t(R)$ in $\mathcal{D}'$ is at least $B^{(d'+r'+1)}(R)$ (we do not need the stronger fact that $t$ is strictly monotone in each argument). But $B^{(d'+r'+1)}(R) > S_k$, since $R \in C$. Hence, $t(R) > S_k$. This is the contradiction that was to be shown. So indeed, for each object $R \in C$ algorithm $\mathcal{A}$ must use a random access. Hence, $r' \geq |C|$.

Set $d'' = h(|C| + k) + d' + r' + 1$. We now show that CA halts by depth $d''$. There are two cases, depending on whether or not the escape clause in Step 2 of CA (which says "If there is no such object, then do not do a random access on this step") is invoked at some depth $\widehat{d}$ with $d' + r' + 1 \leq \widehat{d} \leq d''$.

*Case 1:* The escape clause of CA is invoked at some depth $\widehat{d}$ with $d' + r' + 1 \leq \widehat{d} \leq d''$. There are two subcases, depending on whether or not $d' + r' + 1 \geq k$.

*Subcase 1:* $d' + r' + 1 \geq k$. Then $\widehat{d} \geq d' + r' + 1 \geq k$. Just as in the second paragraph of the proof of Theorem 6.1, we know that the algorithm CA has seen at least $\widehat{d}$ objects by depth $\widehat{d}$ (this is because by depth $\widehat{d}$ it has made $m\widehat{d}$ sorted accesses, and each object is accessed at most $m$ times under sorted access). If CA had seen strictly more than $\widehat{d}$ objects by depth $\widehat{d}$, then the escape clause would not be invoked. Since the escape clause was invoked, it follows that CA must have seen exactly $\widehat{d}$ objects by depth $\widehat{d}$. By depth $\widehat{d}$, the algorithm CA has made exactly $\widehat{d}m$ sorted accesses. Since CA has seen exactly $\widehat{d}$ objects by depth $\widehat{d}$, and since each object is accessed at most $m$ times under sorted access, it follows that each of the $\widehat{d}$ objects that CA has seen has been seen under sorted access in every one of the $m$ lists. Since $\widehat{d} \geq k$, by depth $\widehat{d}$ there are at least $k$ objects that have been seen under sorted access in every one of the lists. (This situation should sound familiar: it is the stopping rule for FA.) For every object that has been seen, there is no uncertainty about its overall grade (since it has been seen in every list), and so no object that has been seen and is not in the top $k$ list is viable. Since each object that has not been seen has $B^{(\widehat{d})}$ value at most equal to the threshold value at depth $\widehat{d}$, and each member of the top $k$ list has grade at least equal to the threshold value, it follows that no object that has not been seen is viable. So there are no more viable objects outside of the top $k$ list, and CA halts by depth $\widehat{d} \leq d''$, as desired.

*Subcase 2:* $d' + r' + 1 < k$. So algorithm $\mathcal{A}$ sees less than $k$ objects before it halts. If database $\mathcal{D}$ contains more than $k$ objects, then there are two objects $R$ and $R'$ that algorithm $\mathcal{A}$ does not see such that algorithm $\mathcal{A}$ outputs $R$ but not $R'$ as part of the top $k$. But then, since algorithm $\mathcal{A}$ does not have information to distinguish $R$ and $R'$, it must make a mistake on some database (either the database $\mathcal{D}$ or the database obtained from $\mathcal{D}$ by reversing the roles of $R$ and $R'$). So database $\mathcal{D}$ cannot contain more than $k$ objects. Since we are assuming throughout this paper that the number of objects in the database is at least $k$, it follows that $\mathcal{D}$ contains exactly $k$ objects. Therefore, at depth $k$ of algorithm CA, all $k$ objects have been seen under sorted access in every list. Similarly to the proof in Subcase 1, it follows that CA halts at depth $k$. Since $k < d''$, we know that CA halts by depth $d''$, as desired.

*Case 2:* The escape clause of CA is not invoked at any depth $\widehat{d}$ with $d' + r' + 1 \leq \widehat{d} \leq d''$. Recall that CA performs random access on viable objects based on their $B$ values. Until they receive a random access after step $d' + r' + 1$ of CA, the members of $C$ have the highest $B$ values. Therefore, within $h|C|$ steps after reaching depth $d' + r' + 1$ (that is, by step $d' + r' + 1 + h|C|$), all members of $C$ will be randomly accessed. We now argue that the next objects to be accessed in CA will be the $R'_i$'s that are output by $\mathcal{A}$ (unless they have been randomly accessed already.) Here we will appeal to the strict monotonicity in each argument of the aggregation function $t$. For a function $t$ that is strictly monotone in each argument, at each step of CA on a database that satisfies the distinctness property and for every object $R$, if $S(R)$ is missing some fields, then $B_S(R) > t(R)$. Therefore at step $d' + r' + 1 + h|C|$ of CA, for all $R'_i$ whose $t$ value has not been determined we have $B^{(d'+r'+1+h|C|)}(R'_i) > t(R'_i) \geq S_k$. Since no other object with $B^{(d'+r'+1+h|C|)}$ value larger than $S_k$ is left, after at most $hk$ more steps in CA, all of $\{R'_1, R'_2, \ldots, R'_k\}$ with missing fields will be randomly accessed and their $t$ value will be known to CA.

We claim that at step $d''$ of CA there are no more viable objects left: first, $M^{(d'')}_k = S_k$, since all of $\{R'_1, R'_2, \ldots, R'_k\}$ have been accessed (in every field) and each of their $W^{(d'')}$ values equals their $t$ values. Since all other objects $R$ with $B^{(d'')}(R) > S_k$ have been accessed, there are more viable objects left, so CA halts.

We have shown that in both cases, the algorithm CA halts by depth $d''$. Recall that when CA gets to depth $d$ it incurs a cost of at most $2mdc_S$. We showed that CA halts by depth $d'' = h(|C| + k) + d' + r' + 1 \leq h(r' + k) + d' + r' + 1$. Hence, the cost CA incurs is at most $2m(h(r' + k) + d' + r' + 1)c_S$, which is $2m(h(r' + k) + d' + r)c_S$ plus an additive constant of $2mc_S$. Now

$$\begin{aligned}
2m(h(r' + k) + d' + r')c_S &\leq 2m(\frac{c_R}{c_S}(r' + k)c_S + (d' + r')c_S) \\
&= 2m(r'(c_R + c_S) + d'c_S + kc_R) \\
&\leq 2m(r'(2c_R) + d'c_S + kc_R) \text{ since by assumption } c_R \geq c_S \\
&\leq 2m(r'(2c_R) + d'c_S + kr'c_R) \text{ since } r' \geq 1 \text{ (see below)} \\
&= 2md'c_S + (4m + k)r'c_R \\
&\leq (4m + k)(d'c_S + r'c_R)
\end{aligned}$$

Since $d'c_S + r'c_R$ is the middleware cost of $\mathcal{A}$, we get that the optimality ratio of CA is at most $4m + k$.

So we need only show that we may assume $r' \geq 1$. Assume not. Then $\mathcal{A}$ makes no random accesses. Now by Theorem 8.5, NRA is instance optimal compared with algorithms that make no random access, and of course the optimality ratio is independent of $c_R/c_S$. Further, the cost of CA is at most twice that of NRA. So CA is instance optimal compared with algorithms that make no random access, such as $\mathcal{A}$, with optimality ratio independent of $c_R/c_S$. □

In the proof of Theorem 8.9, we showed that under the assumptions of Theorem 8.9 (strict monotonicity in each argument and the distinctness property), the optimality ratio of CA is at most $4m + k$. In Theorem 9.2, we give a lower bound that is linear in $m$, at least for one aggregation function that is strictly monotone in each argument.

The next theorem says that for the function min (which is not strictly monotone in each argument), algorithm CA is instance optimal.

**Theorem 8.10:** *Let **D** be the class of all databases that satisfy the distinctness property. Let **A** be the class of all algorithms that correctly find the top $k$ objects for* min *for every database in **D**. Then CA is instance optimal over **A** and **D**, with optimality ratio independent of $c_R/c_S$.*

**Proof (Sketch):** The proof is similar to the proof of Theorem 8.9, where the key point is that for the function min at every step $d$ of CA there can be at most $m$ different $R$'s with the same $B^{(d)}(R)$ value, since $B^{(d)}(R)$ equals one of the fields of $R$ and the distinctness property assures that there are at most $m$ different fields in *all* lists with the same value (this replaces the use of strict monotonicity in each argument). Therefore at step $d' + r' + 1 + h|C|$ there are at most $m$ objects with $B$ value that equals $S_k$, and there are no objects outside of $\{R'_1, R'_2, \ldots, R'_k\}$ whose $B$ value exceeds $S_k$. Since the $B$ value of each member of $\{R'_1, R'_2, \ldots R'_k\}$ is at least $S_k$, it follows that after $hm$ more steps all of $\{R'_1, R'_2, \ldots, R'_k\}$ will be randomly accessed, so there will be no viable objects left and CA will halt. The rest of the analysis is similar to the proof of Theorem 8.9, except that $hk$ is replaced by $hm$. The net result is an optimality ratio of at most $5m$. □

In the proof of Theorem 8.10, we showed that under the assumptions of Theorem 8.10 (the distinctness property with min as the aggregation function), the optimality ratio of CA is at most $5m$. In Theorem 9.4, we give a lower bound that is linear in $m$.

### 8.4 CA Versus Other Algorithms

In this section, we compare CA against two other algorithms. The first algorithm we compare it against is the *intermittent algorithm*, which does random accesses in the same time order as TA does, but simply delays them, so that it does random accesses every $h = \lfloor c_R/c_S \rfloor$ steps. The second algorithm we compare CA against is TA.

| $L_1$ | $L_2$ | $L_3$ |
|---|---|---|
| $(\cdot, \frac{1}{2} + \frac{h-2}{8h})$ | $(\cdot, \frac{1}{2} + \frac{h-2}{8h})$ | $(\cdot, \frac{1}{2} + \frac{h^2-1}{8h^2})$ |
| $(\cdot, \frac{1}{2} + \frac{h-3}{8h})$ | $(\cdot, \frac{1}{2} + \frac{h-3}{8h})$ | $(\cdot, \frac{1}{2} + \frac{h^2-2}{8h^2})$ |
| $(\cdot, \frac{1}{2} + \frac{h-4}{8h})$ | $(\cdot, \frac{1}{2} + \frac{h-4}{8h})$ | $(\cdot, \frac{1}{2} + \frac{h^2-3}{8h^2})$ |
| ... | ... | ... |
| $(\cdot, \frac{1}{2} + \frac{1}{8h})$ | $(\cdot, \frac{1}{2} + \frac{1}{8h})$ | $(\cdot, \frac{1}{2} + \frac{1}{8h^2})$ |
| $(R, \frac{1}{2})$ | $(R, \frac{1}{2})$ | $(R, \frac{1}{2})$ |
| $(\cdot, \frac{1}{8})$ | $(\cdot, \frac{1}{8})$ | ... |
| ... | ... | ... |

Figure 5: Database about CA versus the intermittent algorithm

**CA versus the intermittent algorithm:** We now consider the choice we made in CA of doing random access to find the fields of the viable object $R$ whose $B^{(d)}$ value is the maximum. We compare its performance with the intermittent algorithm, which we just described. We show a database (see Figure 5) where the intermittent algorithm does much worse than CA. Consider the aggregation function $t$ where $t(x_1, x_2, x_3) = x_1 + x_2 + x_3$. Let $c_R/c_S$ be a large integer. Let $\mathcal{D}$ be a database where the top $h - 2$ locations in $L_1$ and $L_2$ have grades of the form $1/2 + i/(8h)$, for $1 \leq i \leq h - 2$, and where none are matched with each other. Location $h - 1$ in the two lists belong to same object $R$, with grade $1/2$ in both of them. Location $h$ in the two lists both have the grade $1/8$. In $L_3$ the top $h^2 - 1$ locations have grades of the form $1/2 + i/(8h^2)$, for $1 \leq i \leq h^2 - 1$, and in location $h^2$, object $R$ has grade $1/2$. Note that the maximum overall grade (which occurs for the object $R$) is $1\frac{1}{2}$ and that all objects that appear in one of the top $h - 2$ locations in lists $L_1$ and $L_2$ have overall grades that are at most $1\frac{3}{8}$ (this is because each object in the top $h - 2$ locations in $L_1$ has grade at most $5/8$ in $L_1$, grade at most $1/8$ in $L_2$, and grade at most $5/8$ in $L_3$.) At step $h$ in CA we have that $B^{(h)}(R) \geq 1\frac{1}{2}$, whereas for all other objects their $B^{(h)}$ value is at most $1\frac{3}{8}$. Therefore on this database, CA performs $h$ sorted accesses in parallel and a single random access on $R$ and then halts. Its middleware cost is therefore $hc_S + c_R = 2c_R$. The intermittent algorithm, on the other hand, does not give priority to checking $R$, and will first do two random accesses for each of the $h - 2$ objects at the top of each of the three lists. Since we take all of these objects to be distinct, this is $6(h - 2)$ random accesses, with a middleware cost of $6(h - 2)c_R$. So the ratio of the middleware cost of the intermittent algorithm to the middleware cost of CA on this database is at least $3(h - 2)$, which can be arbitrarily large.

In particular, Theorem 8.9 would be false if we were to replace CA by the intermittent algorithm, since this example shows that the optimality ratio of the intermittent algorithm can be arbitrarily large for $h$ arbitrarily large.

**CA versus TA:** It is intriguing to consider the differences between CA and TA, even when $c_R/c_S$ is not large. Intuitively, TA beats CA in terms of sorted accesses, and CA beats TA in terms of random accesses. More precisely, TA never makes more sorted accesses than CA, since TA gathers as much information as it can about every object it encounters under sorted access. On the other hand, if we focus on random accesses, then we see that TA does random access to every field of every object that it sees under sorted access. But CA is more selective about its random accesses. It "stores up" objects that it has seen under sorted access, and then does random access only for the object in its stored-up collection with the best potential.

We now consider other advantages of CA over TA. In the database we presented in comparing CA with the intermittent algorithm, the random access cost of TA is the same as that of the intermittent algorithm. So for this database, the ratio of the middleware cost of TA to the middleware cost of CA is at least $3(h-2)$. This is a manifestation of the dependence of the optimality ratio of TA on $c_R/c_S$ and the independence of the optimality ratio of CA on $c_R/c_S$. Furthermore, the fact that at least under certain assumptions, TA has an optimality ratio that is quadratic in $m$, whereas under certain assumptions, CA has an optimality ratio that is only linear in $m$, is also an indicator of the possible superiority of CA over TA in certain circumstances. This requires further investigation. As an example where it might be interesting to compare CA and TA, let the aggregation function be min, let **D** be the class of all databases that satisfy the distinctness property, and let **A** be the class of all algorithms that correctly find the top $k$ objects for min for every database in **D**. We know that TA and CA are both instance optimal in this scenario (Theorems 6.5 and 8.9), and we know that the optimality ratio of CA is independent of $c_R/c_S$ (Theorem 8.9). What are the precise optimality ratios of TA and CA in this scenario? Which has a better optimality ratio when, say, $c_R = c_S$?

TA has an important advantage over CA. Namely, TA requires very little bookkeeping, whereas, on the face of it, CA requires a great deal of bookkeeping. Thus, in CA, for every sorted access it is necessary to update the $B$ value (the upper bound on the overall grade) for every object where not all of its fields are known. As we discussed in Remark 8.7 for NRA, it would be interesting to develop data structures for CA that would lead to a reasonable amount of bookkeeping. We could then compare CA versus TA in realistic scenarios (both by analysis and simulations).

## 9 Lower Bounds on the Optimality Ratio

In this section, we prove various lower bounds on the optimality ratio, both for deterministic algorithms and for probabilistic algorithms that never make a mistake. Each lower bound corresponds to at least one theorem from earlier in the paper.

The next theorem gives a matching lower bound for the upper bound on the optimality ratio of TA given in the proof of Theorem 6.1, provided the aggregation function is strict. As we noted earlier, this lower bound need not hold if the aggregation function is not strict (for example, for the aggregation function max).

**Theorem 9.1:** *Let $t$ be an arbitrary monotone, strict aggregation function with $m$ arguments. Let **D** be the class of all databases. Let **A** be the class of all algorithms that correctly find the top $k$ answers for $t$ for every database and that do not make wild guesses. There is no deterministic algorithm that is instance optimal over **A** and **D**, with optimality ratio less than $m + m(m-1)c_R/c_S$.*

**Proof:** We assume first that $k = 1$; later, we shall remove this assumption. We restrict our attention to a subfamily $\mathbf{D}'$ of $\mathbf{D}$, by making use of positive parameters $d, \psi, k_1, k_2$ where

1. $d, k_1$, and $k_2$ are integers.

2. $\psi = (dm-1)c_S + (dm-1)(m-1)c_R$.

3. $k_2 > k_1 > \max(d, \psi/c_S)$.

The family **D′** contains every database of the following form. In every list, the top $k_2$ grades are 1, and the remaining grades are 0. No object is in the top $k_1$ of more than one list. There is only one object $T$ that has grade 1 in all of the lists, and it is in the top $d$ of one list. Except for $T$, each object that is in the top $k_1$ of any of the lists has grade 1 in all but one of the lists, and grade 0 in the remaining list. It is easy to see that we can pick $k_1$ and $k_2$ big enough to satisfy our conditions, for a sufficiently large number $N$ of objects.

Let $\mathcal{A}$ be an arbitrary deterministic algorithm in **A**. We now show, by an adversary argument, that the adversary can force $\mathcal{A}$ to have middleware cost at least $\psi$ on some database in **D′**. The idea is that the adversary dynamically adjusts the database as each query comes in from $\mathcal{A}$, in such a way as to evade allowing $\mathcal{A}$ to determine the top element until as late as possible.

Let us say that an object is *high in list* $i$ if it is in the top $d$ of list $i$, and *high* if it is high in some list. Since no object is high in more than one list, there are $dm$ high objects. Assume that $\mathcal{A}$ sees at most $dm - 2$ high objects, and hence does not see at least two high objects $S_1$ and $S_2$. Then the adversary can force the answers that $\mathcal{A}$ receives to be consistent with either $S_1$ or $S_2$ being the top object $T$. This is a contradiction, since $\mathcal{A}$ does not have enough information to halt safely, since it does not know the identity of the top object. So $\mathcal{A}$ must see at least $dm - 1$ high objects. Since $\mathcal{A}$ does not make wild guesses, its sorted access cost is at least $(dm - 1)c_S$. There are two cases.

*Case 1:* Algorithm $\mathcal{A}$ sees some high object under sorted access in a list $j$ where it is not high (and hence below position $k_1$ in list $j$, since no object can be in the top $k_1$ positions in more than one list). Then $\mathcal{A}$ has sorted access cost more than $k_1 c_S > (\psi/c_S)c_S = \psi$, as desired.

*Case 2:* There is no high object that $\mathcal{A}$ sees under sorted access in a list where it is not high. Let us say that a high object $h$ is *fully randomly accessed* if $\mathcal{A}$ does random access to $h$ in each of the lists where it is not high. Whenever $\mathcal{A}$ does random access to a high object in a list where it is not high, then the adversary assures that the first $m - 2$ such random accesses have grade 1, and only the final such random access has grade 0 (this is possible for the adversary to continue until it has done $m - 1$ random accesses for all but one of the high objects). Assume that there are at least two high objects $P_1$ and $P_2$ that are not fully randomly accessed. Then the adversary can force the answers that $\mathcal{A}$ receives to be consistent with either $P_1$ or $P_2$ being the top object $T$. This is a contradiction, since once again, $\mathcal{A}$ does not have enough information to halt safely. So there is at most one high object that is not fully randomly accessed. Since there are $dm$ high objects, it follows that $\mathcal{A}$ must make at least $(dm-1)(m-1)$ random accesses, with a random access cost of $(dm-1)(m-1)c_R$. Hence, the middleware cost of $\mathcal{A}$ is at least $(dm-1)c_S + (dm-1)(m-1)c_R = \psi$, as desired.

So in either case, the middleware cost of algorithm $\mathcal{A}$ on the resulting database is at least $\psi$. However, there is an algorithm in **A** that makes at most $d$ sorted accesses and $m - 1$ random accesses, and so has middleware cost at most $dc_S + (m-1)c_R$. By choosing $d$ sufficiently large, the ratio $\frac{(dm-1)c_S + (dm-1)(m-1)c_R}{dc_S + (m-1)c_R}$ can be made as close as desired to $m + m(m-1)c_R/c_S$. The theorem follows in the case when $k = 1$.

We now describe how to modify the proof in the case when $k > 1$. The idea is that we make $k - 1$ of the top $k$ objects easy to find. We modify the databases given in the proof above by creating $k - 1$ new objects, each with a grade of 1 in every list, and putting them at the top of each of the lists. The simple details are left to the reader. □

In the proof of Theorem 6.5 (which assumes strict monotonicity and the distinctness property), we showed that the optimality ratio of TA is at most $cm^2$, where $c = \max\{c_R/c_S, c_S/c_R\}$. In the next theorem, we give an aggregation function that is strictly monotone such that no deterministic algorithm can have an optimality ratio of less than $\frac{m-2}{2}\frac{c_R}{c_S}$. So in our case of greatest interest, where $c_R \geq c_S$, there is a gap of around a factor of $2m$ in the upper and lower bounds. The aggregation function we use for this result is the function $t$ given by

$$t(x_1, x_2, \ldots, x_m) = \min(x_1 + x_2, x_3, \ldots, x_m) \qquad (5)$$

The reason we made use of the unusual aggregation function in (5) is that in the case of min (or an aggregation function such as average that is strictly monotone in each argument), there is an algorithm (algorithm CA of Section 8.2) with optimality ratio independent of $c_R/c_S$ when we restrict our attention to databases that satisfy the distinctness property. Thus, the negative result of the next theorem does not hold for min or average.

**Theorem 9.2:** *Let the aggregation function $t$ be given by (5) above. Let* **D** *be the class of all databases that satisfy the distinctness property. Let* **A** *be the class of all algorithms that correctly find the top $k$ objects for $t$ for every database in* **D**. *There is no deterministic algorithm that is instance optimal over* **A** *and* **D**, *with optimality ratio less than $\frac{m-2}{2}\frac{c_R}{c_S}$.*

**Proof:** As in the proof of Theorem 9.1, we can assume without loss of generality that $k = 1$. We restrict our attention to a subfamily **D'** of **D**, by making use of positive parameters $d$, $N$, and $\psi$, where

1. $d$ and $N$ are integers.
2. $\psi = (d-1)(m-2)c_R$.
3. $N > \max(d, 4\psi/c_S)$, and $N$ is a multiple of 4.

The family **D'** contains each database of the following form. There are $N$ objects. The top $d$ grades in lists 1 and 2 are of the form $i/(2d+2)$ for $1 \leq i \leq d$, and the object with grade $i/(2d+2)$ in list 1 is the one with the grade $(d+1-i)/(2d+2)$ in list 2. Hence, the $x_1 + x_2$ value of these $d$ objects is $1/2$. The grades in the other lists are of the form $i/N$, for $1 \leq i \leq N$. One of the top $d$ objects in lists 1 and 2 has a grade in the half-closed interval $[\frac{1}{2}, \frac{3}{4})$ in each of the other lists. All the rest of the top $d$ objects in lists 1 and 2 have a grade in the half-closed interval $[\frac{1}{2}, \frac{3}{4})$ in all but one of the other lists, and a grade in the open interval $(0, \frac{1}{2})$ in the remaining list. The top object, which we call $T$, is the unique object whose overall grade is $1/2$. Since $T$ has grade less than 3/4 in lists 3, ..., $m$, it occurs after the first $N/4$ objects in each of these $m-2$ lists. Furthermore, simply based on the grades of the top $d$ objects in lists 1 and 2, it is clear that the top object has grade at most $1/2$.

Let $\mathcal{A}$ be an arbitrary deterministic algorithm in **A**. We now show, by an adversary argument, that the adversary can force $\mathcal{A}$ to have middleware cost at least $\psi$ on some database in **D'**. The idea is that the adversary dynamically adjusts the database as each query comes in from $\mathcal{A}$, in such a way as to evade allowing $\mathcal{A}$ to determine the top element until as late as possible. There are two cases.

*Case 1:* $\mathcal{A}$ does at least $N/4$ sorted accesses. Then the sorted access cost of $\mathcal{A}$ is at least $(N/4)c_S > (\psi/c_S)c_S = \psi$, as desired.

*Case 2:* $\mathcal{A}$ does less than $N/4$ sorted accesses. Let us call the top $d$ objects in lists 1 and 2 *candidates*. Thus, $\mathcal{A}$ does not see any candidate under sorted access in any of the lists $3, \ldots, m$. Let us call a grade that is at least 1/2 *high*, and a grade less than 1/2 *low*. Let us say that a candidate $S$ is *fully randomly accessed* if $\mathcal{A}$ does random access to $S$ in each of the lists $3, \ldots, m$. Whenever $\mathcal{A}$ does random access to a candidate in at least one of lists $3, \ldots, m$, then as long as possible, the adversary assures that the first $m - 3$ random accesses have a high grade, and that only the final random access has a low grade (it is possible for the adversary to continue like this until all but one of the candidates is fully randomly accessed). Assume that there are at least two candidates $P_1$ and $P_2$ that are not fully randomly accessed. Then the adversary can force the answers that $\mathcal{A}$ receives to be consistent with either $P_1$ or $P_2$ being the top object $T$. This is a contradiction, since $\mathcal{A}$ does not have enough information to halt safely. So there is at most one candidate that is not fully randomly accessed.

Since there are at least $d - 1$ candidates that are fully randomly accessed, and hence each have at least $m - 2$ random accesses, the random access cost of $\mathcal{A}$ is at least $(d - 1)(m - 2)c_R$. Hence, the middleware cost of $\mathcal{A}$ is at least $(d - 1)(m - 2)c_R = \psi$, as desired.

So in either case, the middleware cost of algorithm $\mathcal{A}$ on the resulting database is at least $\psi$. However, there is an algorithm in **A** that accesses the top $d$ objects in lists 1 and 2, and then makes a random access to object $T$ in each of lists 3, ..., $m$. Its middleware cost is $2dc_S + (m - 2)c_R$. By choosing $d$ sufficiently large, the ratio $\frac{(d-1)(m-2)c_R}{2dc_S + (m-2)c_R}$ can be made as close as desired to $\frac{m-2}{2}\frac{c_R}{c_S}$. The theorem follows. $\square$

The next theorem is somewhat redundant (except for the fact that it deals with probabilistic algorithms), because of Theorem 9.1. We give it because its proof is simple, and because we generalize the proof in the theorem following it.

**Theorem 9.3:** *Let $t$ be an arbitrary monotone, strict aggregation function with $m$ arguments. Let **D** be the class of all databases. Let **A** be the class of all algorithms that correctly find the top $k$ answers for $t$ for every database and that do not make wild guesses. There is no deterministic algorithm (or even probabilistic algorithm that never makes a mistake) that is instance optimal over **A** and **D**, with optimality ratio less than $m/2$.*

**Proof:** As in the proof of Theorem 9.1, we can assume without loss of generality that $k = 1$. We now define a family of databases, each with $m$ sorted lists. There is a parameter $d$. The top $dm$ values in each of the lists is 1, and all remaining values are 0. There is only one object $T$ that has a value of 1 in more than one of the lists, and this object $T$ has value 1 in all of the lists. Therefore $T$ has overall grade 1, and every other object has overall grade 0. Suppose that $T$ has position $d$ in one of the lists, and position $dm$ in all of the other lists.

Let $\mathcal{A}$ be an arbitrary deterministic algorithm in **A**. Consider the following distribution on databases: each member is as above, and the list where $T$ appears in position $d$ is chosen uniformly at random. It is easy to see that the expected number of sorted accesses under this distribution of algorithm $\mathcal{A}$ is at least $(dm + 1)/2$. Since there must be some database where the number of sorted accesses is at least equal to the expected number of sorted accesses, the number of sorted accesses on this database is at least $(dm + 1)/2$, and so the middleware cost of $\mathcal{A}$ on the resulting database is at least $(dm + 1)c_S/2$. However, there is an algorithm in **A** that makes $d$ sorted accesses and $m - 1$ random accesses, and so has middleware cost $dc_S + (m - 1)c_R$. By choosing $d$ sufficiently large, the ratio $\frac{(dm+1)c_S/2}{dc_S + (m-1)c_R}$ can be made as close as desired to $m/2$. The theorem follows (in the deterministic case).

In the case of probabilistic algorithms that never makes a mistake, we conclude as in the conclusion of the proof of Theorem 6.4. $\square$

In the proof of Theorem 8.10, we showed that under the assumptions of Theorem 8.10 (the distinctness property with min as the aggregation function), the optimality ratio of CA is at most $5m$. The next theorem gives a lower bound that is linear in $m$.

**Theorem 9.4:** *Let **D** be the class of all databases that satisfy the distinctness property. Let **A** be the class of all algorithms that correctly find the top $k$ answers for min for every database. There is no deterministic algorithm (or even probabilistic algorithm that never makes a mistake) that is instance optimal over **A** and **D**, with optimality ratio less than $m/2$.*

**Proof:** The proof is obtained from the proof of Theorem 9.3 by modifying the construction slightly to guarantee that we consider only databases that satisfy the distinctness property. The simple details are left to the reader. □

The next theorem gives a matching lower bound for the upper bound on the optimality ratio of NRA given in the proof of Theorem 8.4, provided the aggregation function is strict.

**Theorem 9.5:** *Let $t$ be an arbitrary monotone, strict aggregation function with $m$ arguments. Let **D** be the class of all databases. Let **A** be the class of all algorithms that correctly find the top $k$ objects for $t$ for every database and that do not make random accesses. There is no deterministic algorithm that is instance optimal over **A** and **D**, with optimality ratio less than $m$.*

**Proof:** As in the proof of Theorem 9.1, we can assume without loss of generality that $k = 1$. We restrict our attention to a subfamily $\mathbf{D}'$ of $\mathbf{D}$, by making use of a positive integer parameter $d$. The family $\mathbf{D}'$ contains every database of the following form.

There are $2m$ special objects $T_1, \ldots, T_m, T'_1, \ldots, T'_m$. There is only one object $T$ in the database with a grade of 1 in every list, and it is one of the $2m$ special objects. Thus, the top object $T$ is one of the special objects. For each $i$, let us refer to list $i$ as the *challenge list* for the special objects $T_i$ and $T'_i$. For each $i$, the top $2m - 2$ objects in list $i$ are precisely the special objects except for $T_i$ and $T'_i$. Thus, no special object is in the top $2m - 2$ of its challenge list, but all of the other special objects are. The top $d$ objects in each list have grade 1, and every remaining object in each list has grade 0. If $T = T_i$ or $T = T'_i$, then $T$ is in position $d$ in list $i$. Thus, the unique top object is at position $d$ in some list. Note that each special object is at or below position $d$ in its challenge list, and exactly one special object (the top object) is at position $d$ in its challenge list.

Let $\mathcal{A}$ be an arbitrary deterministic algorithm in $\mathbf{A}$. We now show, by an adversary argument, that the adversary can force $\mathcal{A}$ to have sorted access cost at least $dm$ on some database in $\mathbf{D}'$. The idea is that the adversary dynamically adjusts the database as each query comes in from $\mathcal{A}$, in such a way as to evade allowing $\mathcal{A}$ to determine the top element until as late as possible.

The first $m-1$ times that algorithm $\mathcal{A}$ reaches position $d$ in a list, the adversary forces $\mathcal{A}$ to encounter some object that is not special in position $d$. Thus, the first time that the adversary allows algorithm $\mathcal{A}$ to encounter a special object after position $2m - 2$ is at position $d$ of the last list $i$ that it accesses to depth $d$. Only at that time does the adversary allow the algorithm to discover which of $T_i$ or $T'_i$ is the top object.

It is clear that the sorted access cost of $\mathcal{A}$ on this resulting database is at least $dm$. However, there is an algorithm in $\mathbf{A}$ that makes at most $d$ sorted accesses to one list and $2m - 2$ sorted accesses to each of the remaining lists, for a total of at most $d + (m-1)(2m-2)$ sorted accesses. and so has middleware cost at most $(d + (m-1)(2m-2))c_S$. By choosing $d$ sufficiently large, the ratio $\frac{dmc_S}{(d+(m-1)(2m-2)c_S}$ can be made as close as desired to $m$. The theorem follows. $\square$

| **A** \ **D** $t$ | Every $\mathcal{D}$; Every $t$ | Ref. | Distinctness; $t$ SM | Ref. | Distinctness; $t$ SMV or min | Ref. |
|---|---|---|---|---|---|---|
| Every correct $\mathcal{A}$ (wild guesses ok) | No instance optimal algorithm possible | Thm 6.4 | TA: $cm^2$ | Thm 6.5 | CA: $4m + k$ ; $5m$ for min | Thm 8.9 Thm 8.10 |
| | | | Lower bound: $\frac{m-2}{2}\frac{c_R}{c_S}$ (certain $t$) | Thm 9.2 | Lower bound: $\frac{m}{2}$ | Thm 9.4 |
| No wild guesses | TA: $m + \frac{(m-1)m}{2}\frac{c_R}{c_S}$ | Thm 6.1 | | | | |
| | Lower bound: $m + \frac{(m-1)m}{2}\frac{c_R}{c_S}$ ($t$ strict) | Thm 9.1 | | | | |
| No random access | NRA: $m$ | Thm 8.5 | | | | |
| | Lower bound: $m$ ($t$ strict) | Thm 9.5 | | | | |

Table 1: Summary of Upper and Lower Bounds

## 9.1 Summary of upper and lower bounds

Table 1 summarizes our upper and lower bounds. The rows correspond to the different restrictions on the set **A** of algorithms, and the columns to the restrictions on the set **D** of databases and on the aggregation function $t$. Note that "SM" means "strictly monotone" and "SMV" means "strictly monotone in each variable." "Distinctness" means that **D** is the collection of databases that satisfy the distinctness property. Note also that $c = \max\{\frac{c_R}{c_S}, \frac{c_S}{c_R}\}$. For each such combination we provide our upper and lower bounds, along with the theorem where these bounds are proven. The upper bounds are stated above the single separating lines and the lower bounds are below them. (The upper bounds are stated explicitly after the proofs of the referenced theorems.) The lower bounds may be deterministic or probabilistic.

## 10 Related Work

Nepal and Ramakrishna [**?**] define an algorithm that is equivalent to TA. Their notion of optimality is weaker than ours. Further, they make an assumption that is essentially equivalent to the aggregation function being the min.[16]

Güntzer, Balke, and Kiessling [?] also define an algorithm that is equivalent to TA. They call this algorithm "Quick-Combine (basic version)" to distinguish it from their algorithm of interest, which they call "Quick-Combine". The difference between these two algorithms is that Quick-Combine provides a heuristic rule that determines which sorted list $L_i$ to do the next sorted access on. The intuitive idea is that they wish to speed up TA by taking advantage of skewed distributions of grades.[17] They make no claims of optimality. Instead, they do extensive simulations to compare Quick-Combine against FA (but they do not compare Quick-Combine against TA).

We feel that it is an interesting problem to find good heuristics as to which list should be accessed next under sorted access. Such heuristics can potentially lead to some speedup of TA (but the number of sorted accesses can decrease by a factor of at most $m$, the number of lists). Unfortunately, there are several problems with the heuristic used by Quick-Combine. The first problem is that it involves a partial derivative, which is not defined for certain aggregation functions (such as min). Even more seriously, it is easy to find a family of examples that shows that as a result of using the heuristic, Quick-Combine is not instance optimal. We note that heuristics that modify TA by deciding which list should be accessed next under sorted access can be forced to be instance optimal simply by insuring that each list is accessed under sorted access at least every $u$ steps, for some constant $u$.

In another paper, Güntzer, Balke, and Kiessling [?] consider the situation where random accesses are impossible. Once again, they define a basic algorithm, called "Stream-Combine (basic version)" and a modified algorithm ("Stream-Combine") that incorporates a heuristic rule that tells which sorted list $L_i$ to do a sorted access on next. Neither version of Stream-Combine is instance optimal. The reason that the basic version of Stream-Combine is not instance optimal is that it considers only upper bounds on overall grades of objects, unlike our algorithm NRA, which considers both upper and lower bounds. They require that the top $k$ objects be given with their grades (whereas as we discussed, we do not require the grades to be given in the case where random accesses are impossible). Their algorithm cannot say that an object is in the top $k$ unless that object has been seen in every sorted list. Note that there are monotone aggregation functions (such as max, or more interestingly, median) where it is possible to determine the overall grade of an object without knowing its grade in each sorted list.

Natsev et al. [?] note that the scenario we have been studying can be thought of as taking joins over sorted lists where the join is over a unique record ID present in all the sorted lists. They generalize by considering arbitrary joins.

## 11 Conclusions and Open Problems

We studied the elegant and remarkably simple algorithm TA, as well as algorithms for the scenario where random access is impossible or expensive relative to sorted access (NRA and CA). To study these algorithms, we introduced the instance optimality framework in the context of aggregation algorithms, and provided both positive and negative results. This framework is appropriate for analyzing and comparing the performance of algorithms, and provides a very strong notion of optimality. We also considered approximation algorithms, and provided positive and negative results about instance optimality there as well.

---

[16] The assumption that Nepal and Ramakrishna make is that the aggregation function $t$ satisfies the *lower bounding property*. This property says that whenever there is some $i$ such that $x_i \leq x'_j$ for every $j$, then $t(x_1, \ldots, x_m) \leq t(x'_1, \ldots, x'_m)$. It is not hard to see that if an aggregation function $t$ satisfies the lower bounding property, then $t(x_1, \ldots, x_m) = f(\min\{x_1, \ldots, x_m\})$, where $f(x) = t(x, \ldots, x)$. Note in particular that under the natural assumption that $t(x, \ldots, x) = x$, so that $f(x) = x$, we have $t(x_1, \ldots, x_m) = \min\{x_1, \ldots, x_m\}$.

[17]They make the claim that the optimality results proven in [?] about FA do not hold for a skewed distribution of grades, but only for a uniform distribution. This claim is incorrect: the only probabilistic assumption in [?] is that the orderings given by the sorted lists are probabilistically independent.

**Open problems:** Let us say that an algorithm is *tightly instance optimal* (over **A** and **D**) if it is instance optimal (over **A** and **D**) and if its optimality ratio is best possible. Thus, Corollary 8.6 says that NRA is tightly instance optimal, and Corollary 6.2 says that in the case of no wild guesses and a strict aggregation function, TA is tightly instance optimal. In the case of no wild guesses, for which aggregation functions is TA tightly instance optimal?[18] What are the possible optimality ratios? For the other cases where we showed instance optimality of one of our algorithms (as shown in Table 1), is the algorithm in question in fact tightly instance optimal? For cases where our algorithms might turn out not to be tightly instance optimal, what other algorithms are tightly instance optimal?

There are several other interesting lines of investigation. One is to find other scenarios where instance optimality can yield meaningful results. Another is to find other applications of our algorithms, such as in information retrieval. We already mentioned (Remark 8.7 and Section 8.4) the issue of finding efficient data structures for NRA and CA in cases of interest, and of comparing CA versus TA.

## 12 Acknowledgments

We are grateful to Michael Franklin for discussions that led to this research, to Miklos Ajtai, Allan Borodin, Erik Demaine, David Johnson, Madhu Sudan, Andrew Tomkins and Mihalis Yannakakis for useful suggestions, and to Larry Stockmeyer for helpful comments that improved readability.

---

[18]As noted earlier, when the aggregation function $t$ is max, which is not strict, TA is tightly instance optimal, with optimality ratio $m$. Similarly, when $t$ is a constant, TA is tightly instance optimal, with optimality ratio 1. There are aggregation functions where TA is not tightly instance optimal. For example define $t$ by letting $t(x_1, \ldots, x_m) = \min(x_1, x_2)$. It is not hard to see that TA is not tightly instance optimal for this choice of $t$ when $m \geq 3$.